\newcommand{\be}{\begin{equation}}
\newcommand{\ee}{\end{equation}}
\newcommand{\bea}{\begin{eqnarray}}
\newcommand{\eea}{\end{eqnarray}}
\newcommand{\nn}{\nonumber}

\newcommand\s{\scriptscriptstyle}
\newcommand{\g}[1]{(\ref{#1})}
\newcommand{\p}{\s ++}
\newcommand{\m}{\s --}
\newcommand{\A}{\s A}

\newcommand{\1}{\s 1}
\newcommand{\2}{\s 2}
\newcommand{\3}{\s 3}

\newcommand{\Dp}{D^{\p}}
\newcommand{\Dm}{D^{\m}}
\newcommand{\D}{{\cal D}^{\p}}

\def\bo{{\raise-.5ex\hbox{\large$\Box$}}}       
\def\fracmm#1#2{{{#1}\over{#2}}}                
\def\Bar#1{\overline{#1}}                       
\def\VEV#1{\left\langle #1\right\rangle}        

\def\abs#1{\left| #1\right|}                    
\def\low#1{{\raise -3pt\hbox{${\hskip 0.75pt}\!_{#1}$}}}
\def\caja{\mathsurround=0pt}
\def\eqalign#1{\,\vcenter{\openup2\jot \caja
        \ialign{\strut \hfil$\displaystyle{##}$&$
        \displaystyle{{}##}$\hfil\crcr#1\crcr}}\,}

\documentstyle[epsf,12pt]{article}
\begin{document}
\begin{flushright}
DESY 97 - 094 \\
ITP-UH-10/97 \\
JINR E2-97-164\\
hep-th/9706078
\end{flushright}

\vglue.2in

\thispagestyle{empty}

\begin{center}
{INDUCED HYPERMULTIPLET SELF-INTERACTIONS IN $N=2$ GAUGE THEORIES}\\

\vglue.2in

E. A. Ivanov\\ 
\vspace{0.3cm}
{\it Bogoliubov Laboratory of Theoretical Physics, Joint Institute for
Nuclear Research, 141980 Dubna, Moscow Region, Russia; 
eivanov@thsun1.jinr.dubna.su}\\
\vspace{0.3cm}
S. V. Ketov~\footnote{ On leave of absence
from: High Current Electronics Institute of the Russian Academy of Sciences,
\newline ${~~~~~}$ Siberian Branch, pr. Akademichesky 4, Tomsk 634055, 
Russia}\\
\vspace{0.3cm}
{\it Institut f\"ur Theoretische Physik, Universit\"at Hannover,
Appelstr.2, 30167, Hannover, Germany; ketov@itp.uni-hannover.de}\\
\vspace{0.3cm}
and \\
\vspace{0.3cm}
B. M. Zupnik~\footnote{On leave of absence from: Institute of Applied Physics,
Tashkent State University, Tashkent 700095, \newline ${~~~~~}$ Uzbekistan} \\

\vspace{0.3cm}
{\it Bogoliubov Laboratory of Theoretical Physics, Joint Institute for
Nuclear Research, 141980 Dubna, Moscow Region, Russia; 
zupnik@thsun1.jinr.dubna.su}
   \end{center}

\vglue.2in
 
\begin{abstract}

We calculate the charged hypermultiplet low-energy effective action in the 
Coulomb branch of the $4D$, $N=2$ gauge theory, by using the harmonic 
superspace approach. We find that the unique leading contribution is given 
by the harmonic-analytic Lagrangian of the fourth order in the hypermultiplet 
superfields, with the induced coupling constant being proportional to central 
charges of $N=2$ SUSY algebra. The central charges are identified with Cartan 
generators of the internal symmetry, and they give BPS masses to the 
hypermultiplets. The induced hyper-K\"ahler metrics appear to be the Taub-NUT 
metric or its higher-dimensional generalizations. Simultaneously, non-trivial 
scalar potentials are produced. Within the harmonic superspace method, we 
show an equivalence between the two known approaches to the $4D$ central 
charges. In the first one, the central charges are obtained via the 
Scherk-Schwarz dimensional reduction from six dimensions, whereas in the 
second one they are generated by a covariantly-constant background gauge 
superfield. Our analysis is extended to a more general situation with 
non-vanishing Fayet-Iliopoulos (FI) terms. The perturbatively-induced Taub-NUT
self-coupling in the Coulomb branch is found to be stable against adding the 
FI term, whereas the non-perturbatively generated (Eguchi-Hanson-type) 
hypermultiplet self-interaction is proposed for the Higgs branch.
\end{abstract}

\newpage

\section{Introduction}

\hspace{0.5cm}
The $N=2$ supersymmetric gauge theories in four dimensions ($4D$) are 
known to possess quite remarkable duality and holomorphy 
properties~\cite{SW}. It is therefore of high importance to understand 
in full 
the structure of quantum effective action of these theories, both 
in the pure 
$N=2$ super-Yang-Mills (SYM) and the matter hypermultiplets sectors. 
In order 
to manifestly preserve all the basic symmetries including $N=2$ supersymmetry
(SUSY) or, at least, be able to fully control their breaking, it is 
natural to 
use the harmonic superspace (HSS) formalism~\cite{GIK1,GI2}. The HSS 
approach 
is the only known one that provides a manifestly $N=2$ supersymmetric 
off-shell description of both $N=2$ gauge multiplets and hypermultiplets. 

The HSS effective action of $N=2$ SYM theory in the sector of gauge fields  
was a subject of study in recent papers~\cite{BBIKO,BBKO}.
The harmonic approach makes it evident that the holomorphic contributions to 
the one-loop effective action (in the Coulomb branch) entirely appear due to 
the non-vanishing central charges in $N=2$ SUSY algebra. 
The central charges are chosen in the form of linear combinations of the  
gauge group Cartan generators, in accordance with the celebrated 
Scherk-Schwarz mechanism~\cite{SS}, and they produce masses for matter 
hypermultiplets. In the context of HSS, such mechanism of generating 
hypermultiplet masses and non-trivial scalar potentials was first 
considered in ref.~\cite{GIK1}. The $N=2$ central charges can be
equivalently treated as a non-trivial constant $N=2$ SYM background, and
this interpretation can also be made transparent within the HSS approach. 

It was suggested in ref.~\cite{Ke} that the same central charges 
also play an important role in the charged matter hypermultiplet sector of 
the $N=2$ HSS effective action (in the Coulomb branch), 
giving rise to non-trivial quantum corrections to 
the free hypermultiplet action and, hence, the non-trivial 
induced hyper-K\"ahler metrics in the physical bosonic 
sector~\footnote{In the Higgs branch, where the
gauge symmetry is completely broken, the hypermultiplet low-energy effective
\newline ${~~~~~}$ action receives no quantum corrections~\cite{BF,APS}.}.
 The basic goal of the present paper is to give 
a correct proof of that proposal, by using quantum HSS computations with the 
massive $q$-hypermultiplet propagator, the mass being induced by the $U(1)$ 
central charges. We compute the one-loop effective action for 
hypermultiplets, 
and  show that the only correction is of the fourth-order in the 
hypermultiplet superfields, with the induced coupling constant  
proportional to the BPS mass parameter. In the case of one hypermultiplet 
we get the familiar Taub-NUT metric~\cite{GIOS}, while for a 
few hypermultiplets we 
find its obvious higher-dimensional generalizations. All the effective 
self-couplings and the corresponding metrics have the 
$SU(2)\times U(1)$ isometry, with the $SU(2)$ 
factor being the $SU(2)_A$ automorphism group of the $N=2$ SUSY 
algebra, and the $U(1)$ factor proportional to the central charge. 
The quartic hypermultiplet self-interaction is the only one that respects 
the $SU(2)_A\times U(1)$ global symmetry (in the case of several matter
hypermultiplets, more $U(1)$ factors can be present). Since
this symmetry is non-anomalous, it protects the quartic form of the analytic 
correction to any order of perturbation theory. 
We also discuss a possibility to gain some other non-trivial 
HSS hypermultiplet actions, e.g. the Eguchi-Hanson action~\cite{GIOT}, 
by using some non-perturbative solutions (torons) to the $N=2$ SYM theory.

Our another purpose is to summarize different approaches to the  
central charges in the HSS formalism and establish their equivalence.

In sects.~2 and 3 we give an account of the basic elements of the 
quantum theory of $N=2$ SYM and 
hypermultiplet harmonic superfields in the presence of central charges, and
discuss the background interpretation of the central charges. As an 
introduction, we remind the reader some results of ref.~\cite{Z1}, where 
the general HSS construction of $N=2$ supersymmetric $4D$ gauge theories 
with a complex central 
charge (including massive propagators for hypermultiplets) was given 
by using 
a dimensional reduction $(DR)$ from six dimensions. It can be considered 
as the 
superspace version of the standard $DR$ method that was proposed within the 
component approach to supersymmetric gauge theories~\cite{SS,BSS,Fa}. 
In this 
method the Cartan generators of spontaneously broken gauge symmetry are 
identified with extra components of the six-dimensional momentum operator, 
which become the central charges of $N=2$ SUSY algebra in four dimensions.
We also discuss another, basically equivalent interpretation of the $U(1)$ 
central charges when they are induced by a non-trivial covariantly-constant 
$N=2$ SYM background. This background triggers spontaneous breaking of the 
gauge symmetry down to its Cartan subalgebra, and thus produces masses for 
the matter hypermultiplets and (via Higgs phenomenon) for the coset gauge 
superfields as
well. It is sometimes advantageous to use that point of view, since it 
gives us an opportunity to construct the {\it massive} hypermultiplet 
propagators as the 
manifestly analytic Green functions of massless hypermultiplets minimally 
coupled to a special (`frozen') external $N=2$ SYM gauge superfield. 

In sect.~4, by using the correct massive hypermultiplet propagator, 
we calculate the one-loop contribution to the hypermultiplet 
effective action in an $N=2$ supersymmetric gauge theory with spontaneously 
broken gauge symmetry. 
A crucial feature of our calculation is the preservation of manifest 
$N=2$ SUSY with central charges at each step. The unique correction is 
of the fourth-order in $q$-hypermultiplets and, according to 
ref.~\cite{GIOS}, it corresponds to the Taub-NUT hyper-K\"ahler metric 
(for a single $q$ -hypermultiplet) or its higher-dimensional 
generalizations (for a few $q$-hypermultiplets). Our approach therefore
 provides the dynamical quantum mechanism of generating the non-vanishing 
Taub-NUT effective coupling constant.
 Since the effective Taub-NUT coupling vanishes in the zero-mass limit,
it is a presence of the non-vanishing $U(1)$ central charges in the theory 
that allows one to get a non-trivial analytic correction. Apart from 
generating the Taub-NUT metric, it is yet another important effect of the 
central charges that they induce a non-trivial scalar potential for the 
hypermultiplets.

In sect.~5 we consider the coupled system of an $N=2$ vector multiplet and a 
hypermultiplet when the additional Fayet-Iliopoulos (FI) term is added. This 
model can be treated along the lines of the preceding sections when one 
chooses the background with a constant scalar field and some constant 
auxiliary components. We calculate the hypermultiplet Green function in 
this background, and argue that the induced (Taub-NUT) terms in the 
hypermultiplet effective action are not affected by adding the FI term. 
The subtle issues of the classical vacuum structure in a presence of the FI 
terms (like spontaneous breaking of $N=2$ supersymmetry versus breaking of 
the gauge symmetry, etc.) are beyond the scope of this paper. Finally, in 
sect.~6 we propose another (Eguchi-Hanson) hypermultiplet self-coupling 
that may be non-perturbatively generated in the Higgs branch.

Our basic notation is the same as in refs.~\cite{GIK1,GI2}, so that 
we invite the reader to consult that references for an introduction 
to the HSS. In particular, we use the following definition for the 
product of four $u^+$ projections of spinor covariant derivatives:
\be
(D^+)^4= \frac{1}{16}(D^+)^2(\bar{D}^+)^2=
\frac{1}{16}(D^{\alpha+}D^ +_\alpha)
(\bar{D}^+_{\dot{\beta}}\bar{D}^{\dot{\beta}+})~,
\label{A1}
\ee
and similarly for the products of Grassmann superspace coordinates:
\be                                                    
(\theta^\pm)^2=\theta^{\alpha\pm}\theta^\pm_\alpha~,\quad
(\bar{\theta}^\pm )^2=\bar{\theta}^\pm_{\dot{\alpha}}
\bar{\theta}^{\dot{\alpha}\pm}~. \label{A2}
\ee
\vglue.2in

\setcounter{equation}{0}
\section{ \label{B} Dimensional reduction, central charges 
and massive \vglue.08in ${~}$ $q$-hypermultiplet propagators }    

\hspace{0.5cm}
In this section we show how the self-consistent HSS formulation 
of $N=2$ SYM theory in four dimensions ($SYM^2_4$) with  central 
charges $Z,\bar{Z}$ can be obtained by the
dimensional reduction from the HSS formulation of $N=1$ SYM theory in 
six dimensions ($SYM^1_6$).    

The classical HSS action of the $SYM^1_6$ theory is given in ref.~\cite{Z1}.
The harmonic superspace description of the $SYM^1_6$ theory is quite similar 
to that of the usual $SYM^2_4$ theory~\cite{GIK1,GI2}. In particular, 
the harmonic $SYM^1_6$ Feynman rules can be obtained as a straightforward 
generalization of the four-dimensional rules . 

Given the gauge group $G_{\rm c}$ (`color'), let $(H^1_{\rm c},\ldots, 
H^r_{\rm c})$  be the hermitian matrix generators of its Cartan subgroup 
$C_{\rm c}=\exp H_{\rm c}$.  Let us also assume that 
the hypermultiplet sector of the theory has the additional global 
symmetry group $G_{\rm f}$ (`flavor'), 
with $(H^1_{\rm f},\ldots, H^p_{\rm f})$ being the
hermitian matrix generators of its Cartan subgroup 
$C_{\rm f}=\exp H_{\rm f}$. 
Without entering into details, we simply state that the basic objects 
of the $SYM^1_6$ theory appear to be an analytic gauge superfield 
$V^{\p}_{(6)}(\zeta_{(6)})$, that takes values in the adjoint representation 
of $G_{\rm c}$ and is inert under the action of $G_{\rm f}$, and analytic 
massless hypermultiplets  $q^+_{(6)} (\zeta_{(6)})$ transforming in a 
representation of $G_{\rm c}\times G_{\rm f}$. The $6D$, $N=1$ HSS 
is a direct 
product of the ordinary $6D$, $N=1$ superspace $(x^m, \theta^i_a)$
($m=1,...6;\; i =1,2;\; a = \alpha, \dot \alpha$) and the internal harmonic 
two-sphere $S^2 \sim SU(2)/U(1)$ parametrized by the harmonic coordinates
$u^{\pm}_i$, $u^{+i}u^-_i=1$. The analytic subspace $\{\zeta_{(6)}\}$ 
of the 
$6D$, $N=1$ HSS has two extra spatial coordinates, $x_{\A}^5$ and 
$x_{\A}^6$, beyond its Grassmann and harmonic coordinates which are common
with the $4D$, $N=2$ HSS. Similarly to their $4D$, $N=2$ 
counterparts, the $SYM^1_6$ harmonic superfields can 
be equally considered either in the central coordinates of 
$6D$, $N=1$ HSS  or
in the analytic coordinates where the analyticity of $V^{\p}_{\s(6)}$ and 
$q^+_{\s(6)}$ is manifest. In what follows, we find convenient 
to combine the 
extra coordinates $x^5$ and $x^6$ (as well as $x^5_A$, $x^6_A$) into 
the complex conjugated ones $z$ and $\bar z$ ($z_{\A}$ and $\bar{z}_{\A}$). 
The relation between the extra coordinates in the central and analytic 
basises of $N=1$, $6D$ HSS then reads
\be 
z_{\A}=z + i(\theta^+ \theta^-), \; \; \bar z_{\A}= \bar z - 
i(\bar \theta^+ \bar \theta^-)~.\label{zaz}
\ee
Their infinitesimal supersymmetry transformations are given by
\be \label{realcova}
\delta z = - i \epsilon^k \theta_k~, \;\; \delta \bar z = 
 i \bar \epsilon^{i}\bar \theta_{i} \;\;\;\; 
(\bar \theta^i_{\dot \alpha} = (\theta_{\alpha \;i})^\dagger )\;,
\ee
and 
\be
 \label{realman}
\delta z_{\A} = 2i \epsilon^i\theta^+u^-_i~, \;\: \delta \bar z_{\A} 
=-2i \bar \epsilon^i \bar \theta^+u^-_i\;.
\ee
It is worth noticing that $z_{\A}$ and $\bar z_{\A}$ are conjugated 
in the sense of the generalized involution of ref.~\cite{GIK1}. In what
follows, we mean by complex conjugation just that.

For a later use, 
we also quote here the structure relations of $6D$, $N=1$ Poincar\'e 
superalgebra, 
\be  
\{Q^i_a, Q^k_b \} = 2\epsilon^{ik} P_{ab}, \quad  P_{ab} = -P_{ab}~,
\label{6dsusy}
\ee
where $a =\alpha, \dot \alpha$~, and 
$$
P_{\alpha \beta} = \epsilon_{\alpha\beta}{\partial \over \partial z}, 
\quad
P_{\dot\alpha \dot\beta} = \epsilon_{\dot\alpha\dot\beta} 
{\partial \over \partial \bar z},  
$$
as well as the corresponding flat spinor derivatives 
\be
D^i_{\alpha\; (6)} = D^i_\alpha - i\theta^i_\alpha 
{\partial\over \partial z}, 
\quad
D_{\dot \alpha i\; (6)} = \bar D_{\dot\alpha i} - 
i\bar\theta_{\dot\alpha\;i} {\partial\over \partial \bar z}~, 
\ee 
where the usual $4D$ spinor derivatives have been introduced,
$D^i_\alpha = \partial /\partial \theta^\alpha_i
+i\bar\theta^{\dot\alpha\;i}\partial_{\alpha\dot\alpha}$ and 
$\bar D_{\dot\alpha\;i} = - \partial /\partial \bar\theta^{\dot\alpha i} 
- i\theta^\alpha_i \partial_{\alpha\dot\alpha}\,$.
  
The dimensional reduction to four dimensions in the HSS framework, 
{\it a l\'a} 
Scherk and Schwarz, assumes a special factorizable dependence of all 
superfields upon the extra space-time coordinates, by identifying 
the translation operators 
along these coordinates with linear combinations of the Cartan generators in 
the full symmetry group $G_{\rm c}\times G_{\rm f}$. It can be done 
in the two 
equivalent ways, either in the central basis of the $6D$, $N=1$ HSS 
or in the analytic one.

In the first approach, one factorizes the extra-coordinate dependence 
in all 
superfields by the use of the universal operator $A(z, \bar z)$,
\be
A(z,\bar z)=\mbox{exp}(z\bar{Z} - \bar{z}Z)~, \label{B1}
\ee
where $Z \in H_{\rm c}\oplus H_{\rm f}$ is, by definition, 
the central-charge 
operator of the dimensionally reduced theory,
\bea
&Z=Z(a) + Z(\mu)~, & \nn \\
& 
Z(a)=\sum\limits^{r}_{n=1} a_n H^n_{\rm c},\quad Z(\mu)= 
\sum\limits^{p}_{l=1}
 \mu_l H^l_{\rm f}~, 
& \label{B2}
\eea
whereas the complex moduli $(a_n, \mu_l)$ define the classical vacuum 
of the theory. 
The operator $Z(\mu)$ does not act on the gauge superfields,
\be
Z(a)\circ V^{\p} \equiv  \sum a_n [H^n_{\rm c}, V^{\p}], \quad 
Z(\mu)\circ V^{\p}=0~.
\ee
The operator $Z(a)$ describes a distribution of central charges {\it 
inside} the irreducible multiplets of the gauge group $G_{\rm c}$. 
On the top 
of that the operator $Z(\mu)$ gives additional contributions to the central 
charge eigenvalues, by acting on the whole matter multiplets.

{}From now on, we mostly deal with the $4D$ formalism. Accordingly, 
we use the 
standard notation and conventions of refs.~\cite{GIK1,GI2} for the 
coordinates 
of the analytic $4D$, $N=2$ superspace,  
$\zeta=(x_{\A}^{\alpha\dot{\beta}},\theta_\alpha^+, 
\bar{\theta}_{\dot{\alpha}}^+, u^{\pm}_i)$.
In the full $4D$, $N=2$ HSS  one can use either 
the analytic basis  $(\zeta,\theta^-_\alpha, 
\bar{\theta}^-_{\dot\alpha} )$ or
the central basis $ (x^{\alpha\dot\beta},\theta_k^{\alpha}, 
\bar{\theta}^{\dot\alpha\; k}, u^{\pm}_i) \equiv (X^M,  u^{\pm}_i) $ 
on equal footing.

The dimensional reduction via the operator $A(z, \bar z)$ defined above 
is universal for both $V^{\p}_{\s (6)}$ and $q^{+}_{\s (6)}$~\cite{Z1},
\be
V^{\p}_{\s (6)}(\zeta_{\s (6)}) = A(z,\bar z)\circ    \hat{V}^{\p}~,\quad 
q^+_{\s (6)}(\zeta_{\s (6)}) = A(z, \bar z)   \hat q^+ ~. \label{symm}
\ee
We now observe that the exact $6D$, $N=1$ analyticity of $V^{\p}_{\s (6)}$ 
and $q^+_{\s (6)}$,  
$$
D^+_{a\;(6)} V^{++}_{\s (6)} = D^+_{a\;(6)} q^+_{\s (6)} = 0~, 
$$ 
amounts to a {\it covariant} $4D$, $N=2$ analyticity for the dimensionally
reduced $4D$ harmonic superfields (denoted by hats),
\be
\hat{D}^+_\alpha \hat V^{\p} = 
\hat{\bar{D}}_{\dot{\alpha}}{}^+ \hat V^{\p} = 0~, \quad 
\hat{D}^+_\alpha \hat q^{+} =  
\hat{\bar{D}}_{\dot{\alpha}}{}^+ \hat q^{+} = 0~,  
\label{B5}
\ee
where we have introduced the notation
\bea
&\hat{D}^+_\alpha=
A^{-1}(z,\bar z)D^+_{\alpha (6)}A(z,\bar z)=   
D^+_\alpha-i\theta^+_\alpha\bar{Z}~, 
\label{B6} \\
& \hat{\bar{D}}_{\dot{\alpha}}{}^+ = 
A^{-1}(z,\bar z)D^+_{\dot\alpha (6)}A(z,\bar z)=
\bar{D}_{\dot{\alpha}}^+ +
i\bar{\theta}^+_{\dot{\alpha}}Z~. & \label{B7}
\eea

In the analytic basis of $4D$, $N=2$ HSS,  
the spinor covariant 
derivatives $D^+$ and $\bar{D}^+$ are just the partial derivatives
\be
D^+_\alpha =\partial/\partial \theta^{\alpha-} \quad {\rm and}\quad
\bar{D}^+_{\dot{\alpha}}=\partial/\partial \bar{\theta}^{\dot{\alpha}-}~.
\label{B8}
\ee
Then the conditions \g{B5} express all the 
terms in the $\theta^-$ expansion of the superfields $\hat V^{\p}$ 
and $\hat q^+$ via their
zeroth-order, exactly-analytic terms. The spinor derivatives 
$\hat D^-_\alpha$
and $\hat D^-_{\dot\alpha}$ follow from their $6D$, $N=1$ counterparts via  
the same $A(z, \bar z)$ transformation as in eqs.~\g{B6} and \g{B7}~, 
\be
\hat{D}^-_\alpha=D^-_\alpha - i\theta^-_\alpha \bar{Z}~,\quad
\hat{\bar{D}}^-_{\dot{\alpha}}=\bar{D}^-_{\dot{\alpha}} + 
i\bar{\theta}^-_{\dot{\alpha}} Z~.
\label{B7b}
\ee 
Here $\hat D^-_\alpha$ and $\hat D^-_{\dot\alpha}$ are the $u^-$ 
projections 
of the standard $4D$, $N=2$ spinor derivatives. 
       
The harmonic derivatives $D^{\pm\pm}$ in the central basis of $6D$,
$N=1$ HSS 
do not contain any derivatives with respect to the space-time coordinates 
and, hence, they commute with $A(z,\bar z)$. When acting on the covariantly 
analytic $4D$ superfields, they can be written either in the central or
analytic basis of $4D$, $N=2$ HSS. In both cases, the central charge 
operators are not involved. For example, in the analytic basis, the $D^{\p}$ 
is given by~\cite{GIK1}
\be
D^{\p}=\partial^{\s++}-2i\theta^{\alpha+}\bar{\theta}^{\dot{\beta}+}
\partial_{\alpha\dot{\beta}} + \theta^{\alpha+}\partial^+_\alpha +
\bar{\theta}^{\dot{\alpha}+}\bar{\partial}^+_{\dot{\alpha}}~, \quad
\partial^{\p} \equiv 
u^{i+}\fracmm{\partial}{\partial u^{i-}}~. 
\ee

Applying the operator $A(z,\bar z)$ to the $6D$, $N=1$ SUSY generators 
$Q^i_{a}$ satisfying the algebra \g{6dsusy} yields the new SUSY 
generators that obey the standard $4D$, $N=2$ SUSY algebra extended by
central charges, 
\be
\hat{Q}^k_{a} =A^{-1}(z,\bar{z})Q^k_{a} A(z,\bar{z})~,~\quad 
\{\hat{Q}^k_a , 
\hat{D}^\pm_b \}=0~,
\quad~[D^{\pm\pm},\hat{Q}^k_a ]=0~.
\label{B9}~
\ee 
Explicitly, these generators are   
\be
\hat{Q}^k_\alpha = Q^k_\alpha + i \theta^k_\alpha \bar Z\;, \;\; 
\hat{\bar Q}_{\dot\alpha k} = \bar Q_{\dot\alpha k} +
i \bar \theta_{\dot\alpha k} Z \;, 
\ee
where the first terms on the r.h.s. are the ordinary $4D$, $N=2$ SUSY 
generators without central charges. We observe that the $N=2$ SUSY 
transformations of an $N=2$ superfield $\Phi$,
\be
\delta \Phi = - (\epsilon^\alpha_k \hat{Q}^k_{\alpha} + 
\bar \epsilon^{\dot\alpha k} \hat{\bar Q}_{\dot\alpha k})\Phi~, 
\ee 
get modified so that the 
supertranslations are now accompanied by the $C_{\rm c} \times C_{\rm f}$ 
rotations with the generators  $\bar Z$ and $Z$ and the group parameters 
$-\delta z$ and 
$\delta \bar z$, respectively, the latter quantities being defined 
in eq.~\g{realcova}. 

Summarizing all the above, we conclude that the role of the similarity 
transformation with the operator $A(z, \bar z)$ consists in 
replacing altogether the translation operators with respect to 
the coordinates $(z, \bar z)$ by the central charges $(\bar Z,Z)$ that 
belong to the Cartan subgroup of the full symmetry group 
$G_{\rm c}\times G_{\rm f}$. The coordinates $(z, \bar z)$ appear 
neither in 
the $6D$, $N=1$ SUSY generators nor in the covariant derivatives 
explicitly, 
whereas the derivatives with respect to these coordinates are only 
present. It is also worth mentioning that eq.~\g{symm}, from the $4D$
 standpoint, is a particular transformation from the global symmetry group.
 Having that in mind, the dimensional reduction 
procedure just described guarantees that all the 
$G_{\rm c}\times G_{\rm f}$ 
invariant (or, at least, $(Z, \bar Z)$-neutral) objects to be 
composed out of 
the $6D$, $N=1$ analytic superfields and the spinor or harmonic derivatives 
acting on them (e.g, the invariant Lagrangians) become the $(z, \bar z)$ 
independent $4D$, $N=2$ HSS quantities. Moreover, any $6D$, $N=1$ analytic 
object of this kind (e.g., the $q^+$ hypermultiplet Lagrangian) goes 
over into 
a $4D$, $N=2$ manifestly analytic (i.e. containing no $\theta^-$ dependence) 
quantity.    

Before explaining how the mass and the correct massive
$q^+$-hypermultiplet HSS propagator come out in this setting, 
let us first describe another, equivalent way 
of dimensional reduction from the $6D$, $N=1$ HSS to the 
centrally-extended 
$4D$, $N=2$ HSS. It is defined in the analytic basis of 
the $6D$, $N=1$ HSS and has the advantage of yielding manifestly analytic 
$4D$, $N=2$ superfields. 

In this alternative version of the $DR$ procedure, instead of the operator
$A(z, \bar z)$ of eq.~\g{B1} one uses the modified operator 
\be   
\tilde{A} (z_{\A}, \bar z_{\A}) = \exp (z_{\A}\bar Z  - \bar z_{\A} Z) = 
 \exp(v) A(z, \bar z)~, \label{anA} 
\ee
where 
\be 
v \equiv i(\theta^+\theta^-)\bar Z +i(\bar \theta^+ \bar \theta^-)Z ~.
\label{B10}         
\ee
The modified $DR$ relations ~\g{symm} take the form   
\be
V^{\p}_{\s (6)}(\zeta_{(6)}) = \tilde{A}(z_{\A},\bar z_{\A})\circ 
\tilde{V}^{\p}~,
\quad 
q^+_{\s (6)}(\zeta_{(6)}) = \tilde{A}(z_{\A}, 
\bar z_{\A}) q^+ ~.  \label{B11}
\ee
The $4D$ covariant derivatives and generators are obtained from their $6D$ 
counterparts by the $\tilde{A}(z_A, \bar z_A)$ rotation according to the
 pattern of eq. \g{B7}. 

It is obvious that the $N=1$ analyticity of the $6D$ superfields on the 
l.h.s. of eq.~\g{B11} implies manifest $N=2$ analyticity of the $4D$ 
superfields 
on the r.h.s of eq.~\g{B11}. Indeed, in the analytic basis, the  $6D$,
$N=1$ derivatives $D^+_a$ are `short', i.e. contain no differentiations 
with respect to the space-time coordinates. Therefore, they commute with
$\tilde{A}(z_{\A}, \bar z_{\A})$ (this statement is, of course, 
basis-independent). 
As a result, we have 
\be
D^+_\alpha \tilde{V}^{\p} = \bar D^{+}_{\dot \alpha} \tilde{V}^{\p} = 0, 
\quad      
D^+_\alpha q^{+} = \bar D^{+}_{\dot \alpha} q^{+} = 0~. \label{analyt}
\ee

On the contrary, the harmonic derivatives do not commute with the
$\tilde{A}(z_{\A}, \bar z_{\A})$, so that they acquire 
additional `connection' terms 
when acting on  $\tilde{V}^{\p}$ and $q^+$,  
\be 
{\cal D}^{\pm\pm} = \tilde{A}^{-1}(z_{\A}, \bar
z_{\A})D^{\pm\pm}\tilde{A}(z_{\A},\bar z_{\A}) = D^{\pm\pm} + 
(D^{\pm\pm}v) \equiv 
D^{\pm\pm} + v^{\pm\pm}~, \label{analD} 
\ee
where we have introduced the central charge `connections' 
\be   
v^{\pm\pm} \equiv i(\theta^{\pm}\theta^{\pm})\bar Z +
i(\bar \theta^{\pm} \bar \theta^{\pm})Z~. \label{defv}  
\ee

Since the two $DR$ operators, 
$A(z,\bar z)$ and $\tilde{A}(z_{\A},\bar z_{\A})$, are 
related via the `{\it bridge}' $e^v$ (see eq.~\g{anA}), it is evident 
that the two ways of the reduction $(6D,~N=1)\rightarrow (4D,~N=2)$ 
are actually 
equivalent. The reduced covariantly analytic and manifestly 
analytic $4D$, $N=2$ 
superfields are related to each other via the bridge $e^v$,
\be
\hat{V}^{\p}=e^v\circ \tilde{V}^{\p}=e^v \tilde{V}^{\p}e^{-v},\quad 
\hat q^+=e^v q^+~. 
\label{c1}
\ee
The similar relation can be established between the 
differential operators in both settings, 
\be   \label{relAA} 
{\cal D} = e^{-v} \hat{D} e^v~, 
\ee   
where the hats refer to the covariantly analytic representation,
$ \hat{D}=\{\hat D^{\pm}_\alpha, \hat{\bar{D}}_{\dot\alpha}{}^{\pm}$, 
$D^{\pm\pm}\}$, whereas the corresponding curved notation refers to the 
manifestly analytic representation, respectively, 
${\cal D}=\{{\cal D}^{\pm}_{\alpha}, 
\bar{\cal D}^{\pm}_{\dot\alpha}$, ${\cal D}^{\pm\pm}\}$. Explicitly, 
the transition relations are 
\bea
{\cal D}^{\pm\pm} &=& e^{-v} D^{\pm\pm} e^{v}, \quad
{\cal D}^-_{\alpha} = 
D^-_{\alpha} -2i \theta^-_\alpha \bar Z\;, \label{relbas1} \\
\bar{\cal D}^-_{\dot{\alpha}} 
&=& \bar{D}^-_{\dot{\alpha}} + 2i 
\bar{\theta}^-_{\dot{\alpha}} Z,\quad
{\cal D}^+_{\alpha} = D^+_\alpha~, \quad
\bar{\cal D}^+_{\dot{\alpha}} = 
\bar{D}^+_{\dot{\alpha}}~. \label{relbas2}       
\eea

The realization of $N=2$ SUSY in the manifestly analytic representation 
can be easily obtained from eqs.~\g{realcova} and ~\g{B9} by 
using the general 
formula \g{relAA}. The corresponding generators in the analytic basis~,
\be
{\cal Q}^k_a =\tilde{A}^{-1} (z_{\A}, \bar z_{\A})Q^k_a
 \tilde{A}(z_{\A}, \bar z_{\A})=e^{-v} \hat{Q}^k_a  e^v ~,
\label{Killing}
\ee
(anti)commute with the covariant derivatives ${\cal D}$.
The associated 
$4D$, $N=2$ SUSY transformations of superfields 
differ from the standard ones (without central charges) by
the $C_{\rm c}\times C_{\rm f}$ rotations with the generators 
$\bar Z, Z$ and the group parameters 
$-\delta z_{\A}, \delta \bar{z}_{\A}$ defined in eq.~\g{realman}. 

Having established the equivalence of both $DR$ procedures, 
we are now in a
position to construct the superfield propagators for $\tilde{V}^{\p}$ and
 $q^+$ in the case of $N=2$ SUSY with central charges. We postpone the 
discussion of $\tilde{V}^{\p}$ until the next sect.~3 and concentrate here 
on the hypermultiplet case.

The free part of the analytic invariant Lagrangian for a 
$4D$-hypermultiplet 
directly follows from the analogous Lagrangian in $6D$, $N=1$ HSS. 
In accordance
with the two versions of the $DR$ procedure, the former   
can be written down in the two equivalent forms, by using either 
covariantly
analytic or manifestly analytic superfields, $\hat q^+$ or $q^+$, 
respectively,
\be
{\cal L}^{(+4)}_{q\;({\rm free})} =  \;\bar{\hat{q}}{}^+ \Dp \hat q^+ = 
 \; \bar q^+ \D q^+~ = 
 \; \bar q^+ (D^{\p} + v^{\p}) q^+~ .\label{qlag}
\ee
Note that the term with $v^{\p}$ in eq.~\g{qlag} breaks the global 
$G_{\rm c}\times G_{\rm f}$ 
symmetry down to the $C_{\rm c}\times C_{\rm f}$ symmetry 
(to be more exact, eq.~\g{qlag} respects the symmetry under 
any subgroup of 
$G_{\rm c}\times G_{\rm f}$ commuting with $Z, \bar Z$).  

The free equations of motion are obtained by varying 
eq.~\g{qlag} with respect 
to $q^+$ and $\bar q^+$, and they read as
\be 
 \D q^+ = 0~, \quad {\rm or} \quad  \Dp \hat q^+ = 0~.
\label{eqsmot}
\ee 
By using the algebra of the centrally-extended covariant derivatives
(e.g., in the covariantly analytic basis),
\be
\{\hat{D}^+_\alpha,\hat{D}^-_\beta\}=2i\varepsilon_{\alpha\beta}\bar{Z}~,   
\quad \{\hat{D}^+_\alpha,\hat{\bar{D}}_{\dot{\beta}}{}^{-}\}=
-2i\partial_{\alpha
\dot{\beta}}~, \quad 
[\Dm, \hat D^+_\alpha] = \hat{D}^-_\alpha~,   
\label{B15} 
\ee
one can derive the identity 
\be
-\,\frac{1}{2}(\hat{D}^+)^4 (\Dm )^2 \hat F =
(\bo  + Z\bar{Z})\hat F \equiv \bo^c \hat F 
 \label{B16}
\ee
valid for any covariantly analytic superfield $\hat F$. 
Applying the identity
\g{B16} to the equation $(D^{\m})^2\hat{q}^+=0$, which is a simple 
consequence of eq.~\g{eqsmot} in its second form, then yields
\be 
(\bo + Z\bar Z)\hat q^+ = (\bo + Z\bar Z) q^+ = 0~. \label{kg}
\ee
Eq.~\g{kg} clearly demonstrates that we are dealing with the massive $q^+$ 
hypermultiplet indeed. 

Because of the existence of the two forms of equations of motion for the
$q^+$-hypermultiplet in eq.~\g{eqsmot}, it is natural to introduce  two 
representations for a hypermultiplet Green function, in the covariantly
analytic frame and in the manifestly analytic one, respectively. The two 
representations turn out to be related via the bridges $e^v$ (see below).

Let us derive the covariantly analytic Green function first. 
To write down the
equation it satisfies, we introduce the covariantly analytic 
$\delta$-function 
\be
\hat{\delta}^{(1,3)}_{v\;ab}(1|2)
=(\hat{D}^+_{\s 1})^4_{ab}(v)\delta^{12}(X_{\s 1}-X_{\s 2})
\delta^{(-3,3)}(u_{\s 1},u_{\s 2})\;, \label{B17} 
\ee
where we have explicitly written down the representation indices 
of $Z$ and 
$\bar Z$, and denoted 
\be
\hat{D}^+_a (v) \equiv e^v D^+_a e^{-v}~.
\ee
The $\delta$-function \g{B17} has many features in common  
with the usual manifestly analytic $\delta$-functions ~\cite{GIK1,GI2}, 
namely,
\be
\hat{\delta}^{(1,3)}_{v\;ab}(1|2) = (\hat{D}^+_{\s 2})^4_{ba}(-v^T) 
\delta^{12}(X_{\s 1}-X_{\s 2})\delta^{(1,-1)}(u_{\s 1},u_{\s 2}) \equiv  
\hat{\delta}_{v\;ba}^{(3,1)}(2|1)~, \label{prop1} 
\ee
and
\bea
(\hat{D}^+_{1 \alpha, \dot \alpha})_{ab}(v)\hat{\delta}_{v\;bc}^{(1,3)}
(1|2) 
&=& 0~, \nn \\
(\hat{D}^+_{2 \alpha, \dot\alpha})_{ab}(-v^T) 
\hat{\delta}_{v\;cb}^{(1,3)}(1|2) & = & 0~,\nn \\
D^+_{2 \alpha, \dot\alpha}\left(\hat{\delta}_v^{(1,3)}(1|2)\hat q^+(2)
\right) &=&  0~. \label{B19}
\eea
Note that $v^T = v$ in a basis where all the Cartan generators 
are diagonal. In what follows we always assume that such basis is chosen. 
Remarkably, it is the {\it flat} spinor derivatives that appear in the 
last line
of eq.~\g{B19}. This fact is a consequence of the important relation 
\be \label{deltarel}
\hat \delta^{(1,3)}_v(1|2) = e^{v_1}
\delta_A^{(1,3)}(1|2) 
e^{-v_2}~, 
\ee
where 
$$
\delta_A^{(1,3)}(1|2) =  
(D^+_{\s 1})^4 \delta^{12}(X_{\s 1}-X_{\s 2})
\delta^{(-3,3)}(u_{\s 1},u_{\s 2}) 
$$ 
is the standard  $4D$, $N=2$ HSS analytic $\delta$-function~\cite{GI2}. 
We are
therefore allowed to use the ordinary analytic measure when 
integrating over
the analytic subspace with the covariantly-analytic $\delta$- function, 
{\it viz.}  
\be
\hat q^+(1)=\int d\zeta_{\s 2}^{(-4)}\hat{\delta}^{(1,3)}_v(1|2)
\hat q^+(2)~.
\label{B20}
\ee

The covariantly-analytic Green function of the massive hypermultiplet
is defined by the equation
\be
\Dp_{\s 1}\hat{G}^{(1,1)}_v(1|2)=\hat{\delta}^{(3,1)}_{v}(1|2)~,
\label{B21}
\ee
and it reads
\be
i \langle \hat{q}^+(1)|\hat{\bar{q}}^+(2) \rangle 
=  \hat{G}^{(1,1)}_v(1|2)=-{1\over \bo^c_1} (\hat{D}^+_{\1})^4(v)
(\hat{D}^+_{\2})^4(-v) \delta^{12}(X_{\s 1}-X_{\s 2})
\frac{1}{(u^+_{\s 1}
u^+_{\s 2})^3}~,
\label{B25}
\ee 
where we have suppressed the internal symmetry indices of $q^+$ and  
$\bar q^+$. In order to verify that the expression \g{B25} satisfies 
eq.~(\ref{B21}) indeed, one should apply the identity 
\be
D^{\p}_{\1}\frac{1}{(u^+_{\s 1}u^+_{\s 2})^3}=\frac{1}{2}(D^{\m}_{\s 1})
^2\delta^{(3,-3)}(u_{\s 1},u_{\s 2})~
\label{B26}
\ee
and then make use of the identity \g{B16} and the representation \g{prop1}. 

In addition, it is easy to check the following important symmetry property:
\be
\hat{G}^{(1,1)}_{-v}(2|1)=-\hat{G}^{(1,1)}_v(1|2)~.
\label{B22}
\ee

The massive $q^+$-hypermultiplet propagator was first obtained 
in ref.~\cite{Z1}
in the covariantly analytic form given above. It can be used for quantum 
perturbative calculations in HSS, e.g., along the lines of 
the massless case 
considered at length in ref.~\cite{GI2}, after obvious modifications of
the Feynman rules there. Sometimes, one finds it more convenient to 
use the manifestly analytic representation for the massive propagator. It 
can be obtained from the covariantly analytic one via the bridges 
$e^v$ and $e^{-v}$, by using the general formulas \g{c1} and \g{relAA}:
\be 
i \langle q^+(1)| \bar{q}^+(2) \rangle 
=
G^{(1,1)}_v(1|2)=\mbox{exp}(v_{\2}-v_{\1})\hat{G}^{(1,1)}_v(1|2)~,
\label{B23}
\ee
or, more explicitly, 
\be
i \langle q^+(1)| \bar{q}^+(2) \rangle =
-\frac{1}{\bo^c_1}(D^+_{\1})^4(D^+_{\2})^4
e^{(v_{\2} - v_{\1})}\delta^{12}(X_{\1}
- X_{\2})\frac{1}{(u^+_{\1}u^+_{\2})^3}~.
\label{B24}
\ee
When applying the operator $\Dp_1$ to the both sides of eq.~\g{B23}  and
making use of eqs.~\g{relbas1} and \g{deltarel}, it is easy to show that 
the defining equation \g{B21} for $\hat{G}^{(1,1)}_v(1|2)$ implies the 
following equation for $G^{(1,1)}_v(1|2)$: 
\be
{\cal D}^{\p}_1 G^{(1,1)}_v(1|2) = \delta_A^{(3,1)}(1|2)~.\label{ge2} 
\ee
Eq.~\g{ge2} agrees with the manifestly analytic form of the equations of
motion for the massive  $q^+$ hypermultiplet in eq.~\g{eqsmot}, 
and it completes the proof.       
\vglue.2in 

\setcounter{equation}{0} 

\section{ \label{C} The background interpretation of central charges}

\hspace{0.5cm}
Our method of introducing central charges for hypermultiplets in $HSS$
does not use the local gauge symmetry. The only relevant property for the
dimensional reduction is the invariance of the action under the global
group $G_{\rm c}\times G_{\rm f}$ or even under its Cartan subgroup
$C_{\rm c}\times C_{\rm f}$. Given this invariance, the central charges 
can be introduced `by hand', directly in the $4D$ setting, without any 
reference to the $DR$ procedure. 

On the other hand, within the standard formalism of $SYM_4^2$ theory there 
exists a natural dynamical mechanism of {\it spontaneous} generation of the 
$U(1)$ central charges. They emerge as a result of the appearance of 
non-zero vacuum expectation value of the gauge superfield $V^{\p}$. 
This interpretation is actually suggested by the HSS formulation of 
$N=2$ gauge theories \cite{GIK1}, and it is discussed in the recent 
preprints \cite{BBIKO,Ku}. 

Let us start from the full $q^+$ hypermultiplet action enjoying 
invariance under both the gauge group $G_{\rm c}$ and the global group 
$G_{\rm f}$~,
\be \label{qV}
S_q = -{1\over \kappa^2} \int d\zeta^{(-4)} \, {\cal L}^{(+4)}_q, \quad 
{\cal L}^{(+4)}_q = \, \bar q^+(D^{\p} + V^{\p}) q^+ ~, 
\ee
and assume that by some dynamical reason (e.g., as a result of 
non-perturbative effects), $V^{\p}$ develops a non-zero vacuum expectation 
value just of the form  \g{defv} and \g{B2}~,
\be \label{separ}
V^{\p} = v^{\p} + \tilde{V}^{\p} = i(\theta^+\theta^+)\bar Z + i(\bar
\theta^+ \bar \theta^+) Z + \tilde{V}^{\p}\;, \;\; \langle \tilde{V}^{\p} 
\rangle = 0\;,
\ee
where only the moduli $a_n$ associated with the Cartan subgroup $C_{\rm c}$ 
are assumed to contribute.
 One could, of course, reproduce the full form 
\g{B2} of $(Z, \bar Z)$ by promoting the global $G_{\rm f}$ symmetry to the 
local one. However, without loss of 
generality, we accept here that the gauge group is $G_{\rm c}$ and, hence, 
it is the $a_n$ parts that are only present in $Z, \bar Z$. The moduli $a_n$ 
then get the meaning of the vacuum expectation values for the physical 
scalar 
field components of the $N=2$ gauge multiplets associated with the Cartan
generators of $G_{\rm c}\,$. We still reserve the right to add `by hand' 
the $\mu_l$ parts associated with the `flavor' Cartan subgroup $C_{\rm f}$.

The non-vanishing background value $v^{\p}$ of $V^{\p}$ 
is a particular 
solution of the superfield equations of motion of the $N=2$ gauge theory 
(we assume that all $q^+$ have vanishing vacuum expectation values)~, 
\be
(D^+)^4 v^{\m}=0,\quad \Dp v^{\m}-\Dm v^{\p}=0~,
\label{C1}
\ee
where $v^{\m}$ is given by eq.~\g{defv} and can be viewed 
as the background gauge connection for the harmonic derivative $D^{\m}$. 
The corresponding 
background $N=2$ chiral superfield strength $w$ amounts to a constant 
scalar component, while all its fermionic, $U(1)$-gauge and 
auxiliary components vanish,
\be \label{C4}
w = {i\over 4}(\bar D^+)^2 v^{\m} = Z\;.
\ee
Accordingly, all the covariant derivatives of $w$ vanish too, so that the 
associated background is covariantly-constant. 

In a generic case the background $v^{\m}$  
spontaneously breaks $G_{\rm c}$ down to its Cartan subgroup~\footnote{For 
some particular values of the moduli $a_n$ in eq.~\g{B2} the vacuum 
stability
group commuting with $(Z,\bar{Z})$ \newline ${~~~~~}$ can be enhanced.}, 
and simultaneously 
generates $U(1)$ central charges in $N=2$ superalgebra by the reasons 
obvious from the previous sect.~2. Indeed, the separation of $v^{\p}$ in the 
$q^+$ Lagrangian $\g{qV}$ according to \g{separ} redefines the free part of 
this Lagrangian, thus making it to coincide with eq.~\g{qlag}. The invariance 
group of the latter is just the $(Z, \bar Z)$-extended $N=2$ SUSY. The full 
algebra of the $(Z, \bar Z)$-extended covariant derivatives is naturally 
recovered as the algebra of the $N=2$ gauge-covariantized derivatives 
in the 
above special background. From this point of view, the covariantly 
analytic and 
manifestly analytic frames introduced in the previous section are 
nothing but, respectively, the $\tau$ and $\lambda$ frames of 
the HSS $N=2$ gauge theory (in the terminology of ref.~\cite{GIK1}) 
adapted to the background $v^{\pm\pm}$. 

It is worthy to notice that the $(Z, \bar Z)$-extended $N=2$ SUSY 
can now be interpreted as the invariance group of the above background: 
the modified supercharges naturally appear as the corresponding 
Killing spinors. 
Indeed, the standard $N=2$ generators 
$Q_\alpha^i$, 
$\bar Q_{\dot\alpha \;i}$ 
have a non-vanishing action on the background 
solution $v^{\p}$,
\be \label{noninv}
\delta_{susy}v^{\p} =  
-(\epsilon Q + \bar \epsilon \bar Q)v^{\p} = 
-2i \epsilon^i\theta^+ u^+_i 
\bar Z -2i \bar \epsilon^i\bar \theta^+ u^+_i Z\;, 
\ee
but this variation can be compensated by the appropriate analytic gauge
transformation 
\be  \label{comp}
\delta_{comp} v^{\p} = D^{\p} \Lambda_{comp}, \;\; 
\Lambda_{comp} = 
2i \epsilon^i\theta^+ u^-_i 
\bar Z + 2i \bar \epsilon^i\bar \theta^+ u^-_i Z~, 
\ee
thus leaving $v^{\p}$ intact. The $N=2$ supersymmetry transformations
of all superfields having non-zero quantum numbers with respect to $(Z, 
\bar Z)$ should then be accompanied by this compensating gauge 
transformation, e.g., 
\be 
\hat{\delta}_{susy} q^+ \equiv (\delta_{susy} + \delta_{comp}) q^{+} = 
-(\epsilon Q + \bar \epsilon \bar Q)q^{+} - \Lambda_{comp}q^+ ~.
\ee
It is a simple exercise to check that the modified $N=2$ SUSY generators 
are just those of the $Z, \bar Z$ extended $N=2$ SUSY in the manifestly 
analytic frame,
\be 
{\cal Q}_\alpha^i = Q_\alpha^i - 2i \theta^+_\alpha u^{-\;i} \bar Z(a)~,
\quad 
\bar{\cal Q}_{\dot\alpha \;i} = \bar Q_{\dot\alpha \;i} - 2i \bar 
\theta^+_{\dot\alpha} u^-_i Z(a)~.
\ee 

It also immediately follows from eqs.~\g{noninv} and \g{comp} that the 
$N=2$ SUSY is not modified on the Cartan $C_{\rm c}$ part of the 
shifted gauge 
superfield $\tilde{V}^{\p}$. On the other hand, the non-diagonal part 
of $\tilde{V}^{\p}$ belonging to the coset $G_{\rm c}/C_{\rm c}$ 
realizes the modified $N=2$ SUSY, as this part possesses non-zero quantum 
numbers with respect to $Z(a), \bar Z(a)$.                

The $(Z, \bar Z)$, $N=2$ supersymmetric modified free part of 
the full $q^+, V^{\p}$ Lagrangian \g{qV},
$$
\, \bar q^+{\cal D}^{\p}q^{+} = 
 \,\bar q^+(D^{\p} + v^{\p})q^{+}\;, 
$$ 
respects invariance only under the unbroken global subgroup $C_{\rm c}$ 
of the full gauge group $G_{\rm c}$. Nevertheless, the full local 
$G_{\rm c}$ invariance is restored in the full action due to the property 
that the $G_{\rm c}$ transformation law of the shifted superfield 
$\tilde{V}^{\p}$ contains an inhomogeneous piece  
under the action of the spontaneously broken $G_{\rm c}/C_{\rm c}$ 
generators. The presence of this piece reflects the fact that one 
of the two $G_c$ algebra-valued real physical scalar fields in 
the $G_{\rm c}/C_{\rm c}$ component of $\tilde{V}^{\p}$ is the 
corresponding Goldstone field which should transform inhomogeneously 
under the broken generators.           

Note that the $(Z, \bar Z)$-extended $N=2$ SUSY 
generators ${\cal Q}$ do not, in general, commute with the 
spontaneously broken generators of $G_{\rm c}$. 
However, it is easy to check that the new 
generators appearing in the commutators merely produce some analytic 
$x$-independent (though $u$- and $\theta$- dependent) $G_{\rm c}$ gauge 
transformations which leave  
the whole action invariant. Therefore, we are allowed 
to say that $N=2$ SUSY 
with the $(Z, \bar Z)$ central charges commutes with $G_{\rm c}$ modulo 
analytic gauge $G_{\rm c}$ transformations.  

The modified free $q^+$ action, as was already mentioned in sect.~2, 
corresponds to the massive $q^+$-hypermultiplet, with the 
BPS mass operator $Z\bar Z$ according to eq.~\g{kg}. 
As is clear from the above consideration, this massive hypermultiplet 
action can be equally treated as the action of the {\it massless} $q^+$ 
in the covariantly-constant  background given by 
$v^{++}$. Accordingly, the massive Green function $G_v^{(1,1)}(1|2)$ of 
eq.~\g{B23} acquires the meaning of the Green function of a  $q^+$ 
hypermultiplet in that special background. It can be considered as 
the particular case of the $q^+$ hypermultiplet Green function 
in the general  $SYM$ background, in the framework of the 
background field formalism for the $SYM^2_4$ theory~\cite{BBKO}. 

Along with the $q^+$-propagator of eq.~\g{B23}, the important 
elements of the 
manifestly analytic Feynman rules in the central-charges-extended 
$N=2$ HSS
are the massive $\omega$-hypermultiplet propagator and the propagator of the 
gauge superfield $\tilde{V}^{\p}$. The first one follows from 
the standard massless $\omega$-propagator of ref.~\cite{GI2} via the same 
modification as in eq.~\g{B23}~,
\be
i \langle \omega (1)| \omega (2) \rangle \equiv G^{(0,0)}_v(1|2) = 
-\frac{1}{\bo^c_1}(D^+_{\1})^4(D^+_{\2})^4
e^{(v_{\2} - v_{\1})}\delta^{12}(X_{\s 1}
- X_{\s 2})\frac{u^-_1u^-_2}{(u^+_{\1}u^+_{\2})^3}~, 
\label{omegapro}
\ee
and satisfies the equation 
\be
({\cal D}^{\p}_1)^2 G^{(0,0)}_v(1|2) = \delta_A^{(4,0)}(1|2)\;.
\ee

The  $\tilde{V}^{\p}$ propagator is almost of the same form as that 
without central charges~\cite{GI2}. In particular, the bilinear term 
in the classical $SYM^2_4$ action for $\tilde{V}^{\p}$ contains the
 projection operator
\be
\Pi_v^{(2,2)}(1|2)=\frac{(D^+_{\1})^4(D^+_{\2})^4\mbox{exp}(v_{\1}-v_{\2})}
{(\bo_{\1}+Z\bar{Z})}\frac{\delta^{12}(X_{\s 1}-X_{\s 2})}
{(u^+_{\1}u^+_{\2})^2}
\label{C11}
\ee
that satisfies the equation
\be
{\cal D}_{\1}^{\p}\Pi_v^{(2,2)}(1|2)=0 \label{C12}~.
\ee
The Green function of $\tilde{V}^{\p}$ in the $N=2$ 
Feynman gauge is given by
\be
i \langle  \tilde{V}^{\p}(1)| \tilde{V}^{\p}(2)\rangle = 
\frac{\delta^{(2,2)}_{\A}(1|2)}
{[\bo_{\1}+Z(a)\bar{Z}(a)]} = \frac{1}{\bo^c_1}(D^+_1)^4 
\delta^{12}(X_1-X_2)\delta^{(-2,2)}(u_1, u_2)~, \label{C13}
\ee
where $Z(a)$ is the central charge operator in the adjoint 
representation of
$G_{\rm c}$. Hence, it is the propagators for the non-diagonal (`charged') 
$G_{\rm c}$ components of the full $\tilde{V}^{\p}$ superfield that are 
actually modified. In the quantum calculations to be given 
in the next sect.~4, 
we will only deal with the diagonal (Cartan) components of $\tilde{V}^{\p}$. 
Their propagators have no mass parameters. 

We end this section with two comments.

It follows from the general theory of non-linear realizations 
\cite{CWZ} that 
in any theory with the spontaneously broken gauge symmetry a non-linear 
redefinition of the field variables exists, after which the broken, 
coset part of the gauge transformations appears to be completely 
hidden and compensated. 
In the unitary gauge (by putting the Goldstone fields to be equal to zero) 
the 
only manifest symmetry of the theory turns out to be the symmetry 
with respect 
to the vacuum stability subgroup, and the only genuine massless gauge fields 
are those associated with this subgroup. The coset gauge fields become 
massive 
(Higgs effect), and they are to be considered on equal footing with the 
matter 
fields. Being applied to our case of the $(q^+, V^{\p})$ system with the 
gauge $G_{\rm c}$ symmetry spontaneously broken down to the $C_{\rm c}$ 
symmetry, this general argument implies that, by a field redefinition, 
one can split the original $(q^+, \tilde{V}^{\p})$ Lagrangian into 
two parts.
The first part is the Lagrangian of the 
$(q^+, \tilde{V}^{\p}_{C})$ system, where the abelian massless gauge 
superfield 
$\tilde{V}^{\p}_{C}$ takes values in the Cartan subalgebra and is minimally 
coupled to $q^+$. The hypermultiplet piece of that Lagrangian is 
\be   \label{min2}
\, \bar q^+({\cal D}^{\p} + \tilde{V}^{\p}_C)q^+ ~.\label{extra}
\ee
The rest of the full Lagrangian contains non-minimal couplings among 
$q^+$ and 
the redefined $G_{\rm c}/C_{\rm c}$ coset parts of the original superfield 
$\tilde{V}^{\p}$, which are massive owing to the superfield Higgs effect. 
This picture is just the HSS interpretation of the Coulomb branch in the 
$SYM^2_4$  theory. Leaving aside the issue of an explicit construction of the
relevant redefinition of the superfields involved, in what
follows we will assume that it exists, and it was already done. In our 
quantum 
calculations we will deal with the abelian superfields $\tilde{V}^{\p}_{C}$ 
and the $(q^+, V^{\p})$ Lagrangian in the form \g{extra}. Contributions 
of the massive modes of the original gauge superfield can be ignored in the 
perturbative effects we are interested in. 

Our last remark is that the vacuum background $v^{\p}$, besides breaking 
the gauge $G_{\rm c}$ symmetry and generating the $(Z, \bar Z$) central 
charges in the $N=2$ SUSY algebra, also gives rise to the 
spontaneous breakdown of another important symmetry. It is the 
global symmetry associated with the $U(1)$ factor of the automorphism 
$U(2)$ symmetry of the original $N=2$ SUSY algebra. It is realized 
as phase rotations of the physical complex scalar field in $V^{\p}$. 
It clearly acquires the Goldstone-type realization after separating 
$v^{\p}$ as in eq.~\g{separ}, with one of the scalar fields of the Cartan 
$\tilde{V}^{\p}_{C}$ as the relevant Goldstone field. Note that the
 low-energy holomorphic Seiberg's correction to the $SYM^2_4$  action was
 originally obtained by integrating the quantum anomaly associated with
 just this $U(1)$ factor~\cite{Seib}. The presence of the same non-zero 
$v^{\p}$ in $V^{\p}$ 
was crucial in a recent derivation  of the Seiberg effective action 
by quantum HSS computations~\cite{BBIKO}. It will be shown in the next 
sect.~4, that 
$v^{\p}\neq 0$  is equally responsible for the non-trivial induced 
self-couplings of the hypermultiplets. It is remarkable that this simple 
mechanism yields so many non-trivial group-theoretical and quantum effects~!
\vglue.2in

\setcounter{equation}{0}
\section{\label{D}  A harmonic-supergraph calculation of the massive 
\vglue.08in
${~~}$ hypermultiplet low-energy effective action }

\hspace{0.5cm}
A few basic examples of harmonic supergraphs with the {\it massless} 
gauge and hypermultiplet superfields were considered in ref.~\cite{GI2}. 
In this 
section, we analyse the recent proposal made by one of us in  ref.~\cite{Ke}
that a {\it massive} hypermultiplet in a spontaneously broken (abelian) 
$N=2$ supersymmetric gauge theory has a non-trivial effective 
self-interaction to be induced by quantum corrections in the one-loop 
perturbation theory.

Let us consider a single charged hypermultiplet $q^+$ for simplicity. 
We find 
convenient to use here a pseudo-real notation for the hypermultiplet,
combining $q^+$ and $\bar q^+$ into a doublet of the so-called
Pauli-G\"ursey $SU(2)$ group,  
$q_{a}^+=(\bar{q}^+,q^+)$, $a,b,\ldots=1,2$, with the central charge being 
realized as
\be
Z_c^b= a (\tau_{\s 3})_c^b~, \label{D1}
\ee
where $a$ is a complex moduli and $\tau_{\s 3}={\rm diag}(1,-1)$ 
is the third
Pauli matrix (we could start from an arbitrary $U(1)$ subgroup of 
$SU(2)_{PG}$ with an arbitrary constant traceless $2\times 2$ 
matrix as the generator and then reduce it to eq.~\g{D1} by a proper 
$SU(2)_{PG}$ rotation). An abelian $N=2$ vector superfield $\tilde{V}^{\p}$ 
also receives a $\tau_{\s 3}$ factor, $(\tilde{V}^{\p})_a^b=
\tilde{V}^{\p}(\tau_{\s 3})_a^b$, so that the 
harmonic covariant derivative of the hypermultiplet superfield reads 
\be
\nabla^{\p} q^{+}_a=D^{\p} q^{+}_a + 
(v^{\p}+\tilde{V}^{\p})_a^b q^{+}_b \equiv
[\D + \tilde{V}^{\p}]^b_a q^+_b ~,
 \label{D2}
\ee
where the central charge operator is represented by the background 
vector superfield $v^{\p}$ (see sects.~2 and 3). 
We introduce the following book-keeping notation for various 
covariantly constant superfields to be relevant for our purposes:
\be
 v^c_b =\nu(\tau_{\s 3})^c_b, \quad (v^{\pm\pm})^c_b=
\nu^{\pm\pm} (\tau_{\s 3})^c_b~,  \label{D3}
\ee
\be
 \nu=i[\bar{a}(\theta^+\theta^-) + a(\bar{\theta}^+\bar{\theta}^-)],\quad
\nu^{\pm\pm}=D^{\pm\pm}\nu~, 
\label{D4}
\ee
\be
{\cal D}^{\pm\pm}=D^{\pm\pm}+\nu^{\pm\pm}\tau_{\3}=e^{-\nu\tau_3}D^{\pm\pm}
e^{\nu\tau_3}~.
\label{D5}
\ee

The $U(1)$ gauge harmonic superfield $\tilde{V}^{\p}$ is considered to be
 massless, 
whereas the hypermultiplet mass squared is equal to $m^2=a\bar{a}=|a|^2$, 
which is just the BPS mass. The extra (Pauli-G\"ursey) global $SU(2)_{PG}$ 
symmetry of the free hypermultiplet action, that rotates $q^{+}_a$ with 
respect to its index
$a$, is broken by the gauge interaction to an abelian subgroup $U(1)$. 
Hence,
the full internal symmetry of the $N=2$ supersymmetric action we start 
with is given by a product of the
global $SU(2)_{\rm A}$ automorphism group and the local $U(1)$ group.

We are going to use in this section the manifestly analytic propagator 
of the 
massive hypermultiplet $q^+$ for our calculations. According to the 
results of sect.~2, it reads
\be
G^{(1,1)}_v(1|2)=-\frac{(D^+_{\1})^4(D^+_{\2})^4}{(\Box_{\1}+a\bar{a})} 
\delta (X_{\1}-X_{\2})
\mbox{exp}[\tau_{\3}(\nu_{\2}-\nu_{\1})]
\frac{1}{(u^+_{\1}u^+_{\2})^3}~.
\label{D6}
\ee

The relevant harmonic supergraph has an internal loop built out of 
two hypermultiplet propagators and two $\tilde{V}^{\p}$ propagators (in 
Feynman gauge the latter are given by eq.~\g{C13} without $Z\bar Z$ in 
the denominator), as well as four external hypermultiplet legs (Fig.~1).
\begin{figure}[b]
\vglue.1in
\makebox{
\epsfxsize=4in
\epsfbox{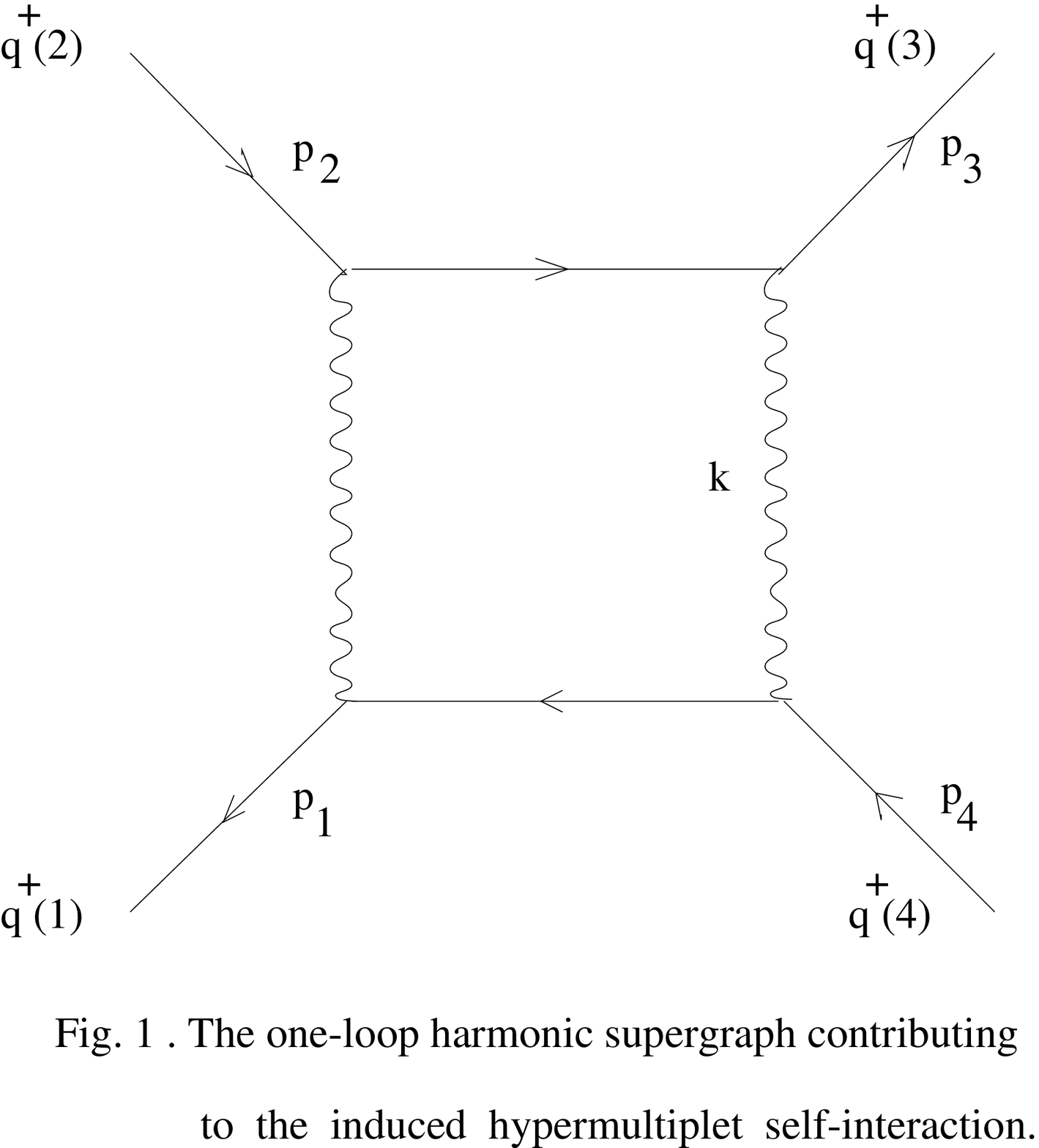}
}
\end{figure}
Note that one of the advantages of our $SU(2)$ notation is 
that we need a single supergraph; it accounts for the two 
inequivalent supergraphs in the $q^+, \bar q^+$ notation. 
 
According to the general strategy of handling such 
supergraphs~\cite{GI2}, one should first restore the {\it full} Grassmann 
integration measures at the vertices. It can be done by taking 
the factors $(D_1^+)^4(D_2^+)^4$ off the hypermultiplet propagators. 
Then one integrates over two sets of Grassmann and harmonic coordinates 
by using the corresponding delta-functions in the integrand. 
There still remain $(D^+)^4$ factors in two $\tilde{V}^{\p}$ propagators.
 After 
integrating by parts with respect to one of these factors, the only 
non-zero contribution comes from the term in which both such factors 
hit one of the two remaining Grassmann $\delta$ functions. Then 
the integral over one more set of the remaining Grassmann 
coordinates can be done by the use of the important identity:
\be
 \delta^8(\theta_{\s 1}-\theta_{\s 2})(D_{\1}^+)^4(D_{\2}^+)^4
\delta^8(\theta_{\s 1}-\theta_{\s 2})=(u^{+i}_1u^+_{2i})^4
\delta^8(\theta_{\s 1}-\theta_{\s 2})~. \label{D7}
\ee
As a result, the single Grassmann integration over $d^8\theta$ remains. 
The supergraph integral then takes the form
\bea
& \Gamma_4= -\,\fracmm{ig^4}{(2\pi)^{16}} \int d^4p_1d^4p_2d^4p_3d^4p_4
 \int d^8\theta \int du_{\s 1}du_{\s 2}\,\fracmm{1}{(u^{+}_{\s 1}
u^+_{\s 2})^2}
\,F(p_1,u_{\s 1}|p_2,u_{\s 2}) F(p_3,u_{\s 1}|p_4,u_{\s 2}) & \nn \\
& \times \int
\fracmm{d^4k \, \delta(p_1-p_2+p_3-p_4)}{k^2(k+p_1-p_4)^2[(k-p_3)^2+m^2]
[(k-p_4)^2+m^2]}~, &   \label{D9}
\eea
where we have introduced the notation $F(p_1,u_1|p_2,u_2)\equiv F(1|2)$
(the $\theta$-dependence is implicit), with
\be
F(1|2)\equiv q^+(1)\exp\{ {\tau}_3[\nu(2)-\nu(1)]\} q^+(2)~. \label{D10}
\ee

According to the definition of the {\it Wilsonian} low-energy 
effective action
(at energies $\ll\Lambda$), we are supposed to integrate over all 
massive fields
as well as over momenta squared $\geq\Lambda^2$ for all the massless fields
present in the fundamental (microscopic) Lagrangian, with the dynamically
generated scale $\Lambda$, $\Lambda<m_{\rm lightest}\neq 0$, 
being an IR-cutoff
simultaneously. In our case, we do not integrate over the massive 
gauge fields 
from the microscopic (non-abelian) Lagrangian. Instead, we simply drop 
them from the very beginning 
since they do not contribute to the low-energy effective 
hypermultiplet self-interaction~\cite{Ke}. Being only interested in 
calculating the leading contribution to the Wilsonian effective action, 
we are also allowed to omit in $\Gamma_4$ any terms that may only 
contribute to higher-order terms in the momentum expansion of the 
effective action.

In order to extract the relevant low-energy contribution out of 
eq.~(\ref{D9}), it is convenient to employ the covariant harmonic 
derivatives $(\D,{\cal D}^{\m})$ defined in eq. \g{D5}. 
This makes integrations by
parts with respect to them simpler since the `bridge' $e^{-v}$ is 
covariantly-constant, 
$\left({\cal D}^{\pm\pm}e^{-v}\right)=0 \Rightarrow 
{\cal D}^{\pm\pm}e^{-v} = e^{-v}D^{\pm\pm}$. The covariant derivatives 
satisfy the algebra
\be
[ {\cal D}^{\p},{\cal D}^{\m} ] = D^{\s 0}~,\label{D11}
\ee
while the flat derivative $D^{\s 0}$ measures the harmonic 
$U(1)$-charge.
 
We insert the relation \g{D11} into the harmonic integral in eq.~(\ref{D9}) 
by identically rewriting 
$$
F(1|2)F(1|2) = {1\over 2}[ {\cal D}_1^{\p}, {\cal D}_1^{\m} ]F(1|2)F(1|2)\;. 
$$  
Then it is easy to argue that only the first term 
$\sim {\cal D}^{\p}{\cal D}^{\m}$ in this identity 
gives the leading low-energy contribution in the local limit. 
Integrating by parts in it, one can cancel the harmonic distribution
$\fracmm{1}{(u^+_{\s 1}u^+_{\s 2})^2}$ by the use of another 
identity~\cite{GI2}
\be
D^{\p}_{\s 1}\fracmm{1}{(u^+_{\s 1}u^+_{\s 2})^2}=D^{\m}_{\s 1}
\delta^{(2,-2)}(u_{\s 1},u_{\s 2})~. \label{D12}
\ee
The harmonic delta-function on the right-hand side of eq.~(\ref{D12})
removes one of the two remaining harmonic integrals, 
and allows us to rewrite eq.~(\ref{D9}) to the form
\be
\Gamma_4\to -ig^4\int \frac{d^4k}{(2\pi)^4}\,\frac{1}{k^4(k^2+m^2)^2}\,S_4~,
\label{D13}
\ee
where we have ignored a dependence upon all the external momenta $p_i$ (the
terms omitted do not contribute to the hypermultiplet low-energy effective
action), and introduced the notation
\be
S_4=\int d^4 xd^8 \theta du\, ({\cal D}^{\m}q^+\cdot q^+)^2~.
\label{D14}
\ee
The $S_4$ is gauge-invariant as it should, and it has the form of the 
{\it full} superspace integral, in formal agreement with the 
non-renormalization `theorem' in superspace. 
However, the connection $v^{\m}$ in
${\cal D}^{\m}$ at $Z\neq 0$ is quadratic in $\theta^-$ because of
eqs.~(\ref{D4}) and (\ref{D5}). In turn, this results in a
$(\theta^-)^4$-dependent term
in the integrand of  $S_4$, and, therefore, an {\it analytic} 
contribution to
the induced hypermultiplet effective action. In other words, it the
non-vanishing central charge $Z$ that is responsible for 
the appearance of 
the leading analytic term in the low-energy effective action, 
already at the 
one-loop level of quantum perturbation theory. The analytic 
part in $S_4$ is 
given by
\be
(S_4)\to\int d^4 xd^8 \theta du\, (\nu^{\m}q^+\tau_{\3}q^+)^2=
-8|a|^2\int d\zeta^{(-4)} (\bar{q}^+ q^+)^2\equiv (S_4)_{\rm analytic}~.
\label{D15}
\ee
The terms omitted, in particular those with
$({\cal D}^{\m})^2 q^+$ or $\D q^+$,
only contribute to higher-order terms in the momentum expansion of the
effective action. Hence, the one-loop induced low-energy hypermultiplet 
self-interaction takes the form~\cite{Ke}
\be
\int d\zeta^{(-4)} {\cal L}^{(+4)}_{\rm ind.~int.}
= \frac{\lambda}{2} \int d\zeta^{(-4)} (\bar{q}^+ q^+)^2~.\label{D16}
\ee
The Lagrangian ${\cal L}^{(+4)}_{\rm ind.~int.}$ is the analytic 
hyper-K\"ahler potential that describes the Taub-NUT metric \cite{GIOS}
(see Appendix). The induced coupling constant 
$\lambda$ appears to be 
\be
\lambda_{\rm 1-loop}=(2g)^4|a|^2\int_{k^2\geq \Lambda^2} 
\frac{d_{\rm E}^4k}{(2\pi)^4} \frac{1}{k^4(k^2+m^2)^2}=
\frac{g^4}{\pi^2}\left[ \frac{1}{m^2}\ln
\left(1+\frac{m^2}{\Lambda^2}\right) -\frac{1}{\Lambda^2+m^2}\right]~,
\label{D17}
\ee
where we have used the BPS mass relation. The induced Taub-NUT 
effective coupling constant $\lambda$ may be renormalized 
in higher orders of 
perturbation theory, unlike the {\it form} of the effective Taub-NUT 
self-interaction that cannot be changed, even non-perturbatively, 
within the 
$N=2$ gauge theory under consideration. The reason for that is the 
non-anomalous global $U(2)=SU(2)_{\rm A}\times U(1)$ symmetry of the 
fundamental action. This symmetry is exact in the full quantum theory. 
The only $SU(2)_{\rm A}\times U(1)$-invariant hypermultiplet self-coupling 
is just the Taub-NUT action. The well-known $U(2)=SU(2)\times U(1)$ 
isometry of the Taub-NUT
metric is just due to the global symmetries of the Taub-NUT hypermultiplet 
action. In this sense, our result is exact and universal.

Similarly, for a spontaneously broken $N=2$ supersymmetric QCD (in the 
Coulomb branch) with vanishing bare hypermultiplet masses, 
one finds a non-trivial 
analytic hyper-K\"ahler potential in the form
\be
{\cal L}^{(+4)}_{\rm ind.~int.}=\frac{\lambda}{2} \sum_{i,j=1}^{N_f} 
\left( \bar{q}^{i+}\cdot q^+_j\right)\left(\bar{q}^{j+}\cdot q^+_i\right)~,
\label{D18}
\ee
where the dots stand for contractions of color indices, and $N_f$ is the
number of flavors. In a more general case with non-vanishing bare 
hypermultiplet masses $m_i$, $i=1,2,\ldots, N_f$, (or, equivalently, with
different eigenvalues of central charge for each matter hypermultiplet),
one finds instead
\be
{\cal L}^{(+4)}_{\rm ind.~int.}=\frac{1}{2}(\bar{q}^+\cdot q^+)\hat{\lambda}
(\bar{q}^+\cdot q^+)~, \label{D19}
\ee
where
\bea
& \hat{\lambda}_{ij}^{\rm 1-loop}
=(2g)^4{\rm Re}[(a+m_i)(\bar{a}+m_j)]
\int_{k^2\geq \Lambda^2}\fracmm{d^4k}{(2\pi)^4}
\fracmm{1}{k^4(k^2+m_i^2+|a|^2)(k^2+m_j^2+|a|^2)}  &  \nn \\
& = \fracmm{g^4{\rm Re}[(a+m_i)(\bar{a}+m_j)]}{2\pi^2(m^2_i
+|a|^2)(m^2_j+|a|^2)} 
\left\{ \ln
\left[(1+\fracmm{m^2_i+|a|^2}{\Lambda^2})(1+\fracmm{m^2_i+|a|^2}{\Lambda^2})
\right] \right. & \\ \nn 
& \left. -\,\fracmm{m^2_i+m^2_j+2|a|^2}{m^2_i-m^2_j} 
\left[ \ln(1+\fracmm{m^2_i+|a|^2}{\Lambda^2})
- \ln(1+\fracmm{m^2_j+|a|^2}{\Lambda^2})\right]\right\}~,~~~m_i\neq m_j & \\ 
\nn
& = \fracmm{g^4 {\rm Re}[(a+m_i)(\bar{a}+m_j)]}{\pi^2(m^2_i+|a|^2)^2}
\left\{ \ln
(1+\fracmm{m^2_i+|a|^2}{\Lambda^2})-\fracmm{m^2_i+|a|^2}{\Lambda^2
+m^2_i+|a|^2}\right\}~,~~~m_i=m_j &  \label{D20}
\eea
Note that $\hat{\lambda}_{ij}^{\rm 1-loop}$ vanishes whenever $a=-m_i$ or
$a=-m_j$.

Summarizing above, we conclude that quantum corrections in an $N=2$ gauge 
theory with the spontaneously broken gauge symmetry produce the effective 
Taub-NUT-type interaction for the charged matter hypermultiplets, which 
preserves $N=2$ supersymmetry, its $SU(2)_A$ automorphisms, 
and the local $U(1)$ rotations. Note that, in the case of several matter
hypermultiplets and, respectively, several independent Cartan generators, 
more non-anomalous $U(1)$ factors can be present in the symmetry group of the 
induced hypermultiplet self-interaction and, hence, in the isometry group of 
the relevant hyper-K\"ahler metric. 

As a simple application, consider the famous Seiberg-Witten model, 
whose initial
(microscopic) action describes the purely gauge $SYM^2_4$ theory, with the 
$SU(2)$ gauge group to be spontaneously broken to its $U(1)$ 
subgroup~\cite{SW}. 
In the non-perturbative region (the Coulomb branch) near a singularity in the 
quantum 
moduli space where a BPS-like (t'Hooft-Polyakov) monopole becomes massless, 
the Seiberg-Witten model is just the {\it dual} $N=2$ supersymmetric QED, 
$V^{++}\rightarrow V^{\p}_D$ and $a\rightarrow a_D$. The t'Hooft-Polyakov
monopole is known~\cite{SW} to belong to a hypermultiplet $q^+_{HP}$ that 
represents the non-perturbative degrees of freedom in the theory. 
Our results 
imply that the induced HP-hypermultiplet 
self-interaction in the vicinity of the 
monopole singularity is regular, and it is given by
\be
 {\cal L}^{(+4)}_{\rm Taub-NUT}(q_{HP})=\fracmm{\lambda_D}{2} 
(\bar{q}^+_{HP}
q^+_{HP})^2~, \label{D21}
\ee
whose induced (one-loop) coupling constant $\lambda_D$ is given by 
eq.~(\ref{D16}) in terms of the dual (magnetic) coupling constant $g_D$ 
and the 
dual moduli $a_D$. Possible non-perturbative generalizations 
of the induced 
hypermultiplet self-interaction from M theory are discussed in sect.~6. 

Another interesting consequence of the induced Taub-NUT hypermultiplet
self-interaction is a non-trivial dynamically generated scalar potential. 
In components, for a single hypermultiplet, it reads 
\be
V(f)=\abs{Z}^2\fracmm{f^i\bar{f}_i}{1+ \lambda f^j\bar{f}_j}~,
 \label{D22}
\ee
where the physical bosonic components $f^i(x)$ of the $q^+$-hypermultiplet
have been defined in Appendix, see eqs.~(A.4) and (A.7). A possibility of
generating non-trivial scalar potentials via non-vanishing 
central charges in
the non-linear, $2D$, $N=4$ supersymmetric sigma-models was noticed earlier 
by Alvarez-Gaum\'e and Freedman in ref.~\cite{AFp}. 
\vglue.2in

\setcounter{equation}{0}
\section{The background induced by Fayet-Iliopoulos terms}

\hspace{0.5cm}
In this section we discuss the modifications due to a presence of the 
additional Fayet-Iliopoulos (FI) terms. First, let us remind the reader that 
we consider the Coulomb branch of the $N=2$ gauge theory  where all the 
gauge symmetries are hidden and compensated  except of the linear one 
associated with the abelian vacuum stability subgroup $C_{\rm c}\,$. 
Of course, the global flavor $G_{\rm f}$ symmetry can survive too. 
We just add the $C_{\rm c}\,$-valued  FI term to the compensated form of 
the action, and we do not care of its possible (dynamical) origin within the 
initial unbroken phase of the theory. We are not going to discuss the 
classical vacuum structure of the $N=2$ vector multiplet/hypermultiplets 
system in the presence of the FI terms (see, however, a brief comment below). 
Our basic aim here is merely to examine how adding the FI terms can affect 
the induced hypermultiplet self-coupling that was found in the previous 
section. 

The HSS superfield form of the FI term is given by  
\be
S_{\s FI}=i\int d\zeta^{(-4)} \, {\rm tr} \,\xi^{\p}\tilde{V}^{\s ++}_C ~, 
\label{E1}
\ee
where $\xi^{\p} \in H$ is the real operator
\be
 \xi^{\p}=\sum_r H^{n}_c \xi^{n\;(kl)} u^+_k u^+_l \equiv 
\xi^{kl} u^+_k u^+_l~,
 \label{E2}
\ee
and the isovectors  $\xi^{n\;(ij)}$ are some constants satisfying 
the reality condition 
$\Bar{\xi^{(ij)}_{(r)}}=\varepsilon_{ik}\varepsilon_{jl}\xi^{(kl)}_{(r)}$, 
and representing the additional moduli of the theory. We use also the
 notation 
$\xi^{\s+-}=\xi^{kl}u^+_k u^-_l $ and $\xi^{\m}=\xi^{kl}u^-_ku^-_l$. 

Adding the term \g{E1} modifies the free part of the $V^{\p}_C$ action and 
we need to diagonalize the latter. This can be accomplished by shifting 
$V^{\p}_C$ as follows 
\be
\tilde{V}^{\p}_C =s^{\p}+V^{\p}_\xi ~,\label{E5}
\ee
with 
\be
s^{\p}\equiv i\xi^{\m}(\theta^+)^2(\bar{\theta}^+)^2~
\ee
(we assume that the part $v^{\p}$ was already separated, being included 
into the free $q$ hypermultiplet action). For $V^{\p}_\xi$ one has the 
standard quadratic action and, hence, the standard form of the 
propagator~\cite{GI2}.

Note that $s^{\p}$ can be interpreted as a new non-trivial background piece 
in the superfield $\tilde{V}^{\p}_C$. 
Indeed, the classical free equation of motion for the abelian potential 
$\tilde{V}^{\p}_C$ now reads 
\be
(D^+)^4 \tilde{V}^{\m}_C=i\xi^{\p}~, \quad {\rm subject~~to}\quad 
{\cal D}^{\p}\tilde{V}^{\m}_C - {\cal D}^{\m}\tilde{V}^{\p}_C = 0~,
 \label{E3}
\ee
and it has a Lorentz-invariant background solution
\be
\langle \tilde{V}^{\p}_C \rangle = s^{\p}~.
\label{E4}
\ee
The FI induced background splitting of the analytic potential formally 
results off shell in a spontaneous $N=2$ supersymmetry breaking since the 
new abelian superfield $V^{\p}_\xi$ has an inhomogeneous $N=2$ supersymmetry 
transformation and hence contains Goldstone fermion components~\footnote
{A modification of the standard $N=2$ algebra in an abelian gauge theory
with spontaneous supersymmetry  
\newline ${~~~~~}$ breaking will be discussed in ref.~\cite{IZ}.}.
However, one can make decisive conclusions about breaking of supersymmetry 
and/or $U(1)$ symmetry only after carefully analyzing the classical vacuum 
on-shell structure of the theory, e.g., along the lines of 
refs.~\cite{Fa2,Ka,ZT}. As was shown there, the structure of the on-shell
vacuum solution for the gauge and hypermultiplet superfields depends on the
choice of initial gauge group and the hypermultiplet interactions. 
The FI-terms produce, in general, spontaneous breaking of  
$U(1)$ gauge symmetries or a simultaneous breaking of $N=2$ supersymmetry 
and $U(1)$ symmetries. In what follows, we shall not consider 
nonvanishing vacuum solutions for the hypermultiplets and restrict 
ourselves to the discussion of the perturbation theory in the 
background \g{E4}. Let us point out once more that this amounts to 
diagonalizing the $V^{\p}_C$ kinetic action in the presence of the
FI term \g{E1}.

The background `bridge'  corresponding to $s^{\p}$ is $\exp(s)$, where  
\be
s={i\over 2} \{ \xi^{\m}(\theta^+\theta^-)(\bar{\theta}^+)^2 +\frac{1}{3}
\xi^{\p}(\theta^-)^2(\bar{\theta}^+\bar{\theta}^-)   
 -\frac{1}{3}\xi^{\s+-}[2(\theta^+\theta^-)(\bar{\theta}^+ \bar{\theta}^-)
+(\theta^+)^2(\bar{\theta}^-)^2]
+ \mbox{c.c.}\}~. 
\label{E6}
\ee
As far as the contribution to the harmonic connection is concerned, we find
\bea
& s^{\m}=\Dm s = {i\over 3}\{ \xi^{\m}[(\theta^-)^2 (\bar{\theta}^+)^2 
+ 2 (\theta^+\theta^-)(\bar\theta^+\bar\theta^-)] + {1\over 2} \xi^{\p}
(\theta^-)^2(\bar{\theta}^-)^2  \nn \\
&- 2 \xi^{\s+-}(\theta^+\theta^-)(\bar{\theta}^-)^2
 +\mbox{c.c.}\}~. &
\label{E7}
\eea

In the analytic basis the spinor derivatives $D^+_a$ are short whereas the 
harmonic and remaining spinor derivatives are given by
\be
{\cal D}^{\pm\pm}_\xi=D^{\pm\pm}+v^{\pm\pm}+s^{\pm\pm} \equiv
 D^{\pm\pm}+v^{\pm\pm}_{\xi}~,
\label{E8}
\ee
\be
{\cal D}^-_{a} = [{\cal D}^{\m}_\xi, D^+_a ] = D^-_a - 
(D^+_a v^{\m}_\xi) ~.\label{E9}
\ee

The basic relation of the modified background geometry reads as usual,
\be
\{\bar{D}^+_{\dot{\alpha}},\bar{\cal D}^-_{\dot{\beta}} \}=
-2i\varepsilon_{\dot{\alpha}\dot{\beta}}Z_{\xi}~.
\label{E10}
\ee
The corresponding chiral background operator (background 
superfield strength) explicitly depends upon the
Grassmann superspace coordinates and the vacuum expectation
values of the scalar and auxiliary fields as follows:
\be
Z_{\xi} = {i\over 4}(\bar D^+)^2v^{\m}_\xi = 
Z + \frac{1}{3}\xi^{kl}(\theta_k\theta_l),\quad
[\partial_{\alpha\dot{\beta}},Z_{\xi} ]=0~.
\label{E11}
\ee
The second part of the operator $Z_{\xi}$ corresponds to 
a special $\theta$-dependent gauge transformation. It is obviously
non-vanishing on any superfield with non-trivial transformation 
properties under the group $C_{\rm c}\,$.

Let us now construct the $q^+$ propagator in the modified background. 
The generalized d'Alambertian $\bo_{\xi}$ for a hypermultiplet $q^+$ 
in the new background is defined by 
\be
-\frac{1}{2}(D^+)^4({\cal D}^{\m}_\xi)^2 F^p = \bo_{\xi}F^p~,
\label{E12}
\ee
where $F^p$ is an analytic superfield of charge $p$ in a
representation of the gauge group $G_{\rm c}$, and the background 
covariant harmonic derivative squared is 
\be
({\cal D}^{\m}_\xi)^2=(D^{\m})^2 + (v^{\m}_\xi)^2 +
 2v^{\m}_\xi D^{\m} + (D^{\m}v^{\m}_\xi)~.
\label{E13}
\ee
It is straightforward to find 
\be
\bo_{\xi}=\bo + Z\bar{Z}
- \frac{i}{3}v^{\p}_\xi \xi^{\m}+ \frac{i}{3}\xi^{\s+-}+A~, \label{E16}
\ee
where $A$ is the differential operator
\be
A=-{i\over 2}(D^{\alpha+}\bar{D}^{\dot{\alpha}+}v^{\m}_\xi)
\partial_{\alpha\dot{\alpha}}
-\frac{i}{2}(D^{+\;\alpha}Z_\xi) D^-_{\alpha} 
-\frac{i}{2}(\bar D_{\dot\alpha}^+\bar Z_\xi)
\bar D^{- \dot\alpha} + {i\over 4}( (D^+)^2 Z_\xi) D^{\m}~,
\label{E17}
\ee
and $Z_\xi, \bar Z_\xi$ are defined by eq. \g{E11}. It is not difficult to 
check the following important properties of the operator $\bo_\xi$~:
\be  
[ D^+_\alpha, \bo_\xi ] = [ \bar D^+_{\dot\alpha}, \bo_\xi ] = 0, 
\ee
\be 
[{\cal D}^{\p}_\xi, \bo_\xi] = -{i\over 2}(D^+Z_\xi)D^+ 
- {i\over 2}(\bar D^+\bar Z_\xi)
\bar D^+ + {i\over 4}((D^+)^2 Z_\xi)(D^0 -1)\;. \label{propbo}
\ee

We are now in a position to construct by analogy with eq.~(\ref{B24}) 
a harmonic propagator for the hypermultiplet in a more general background 
with the spontaneous supersymmetry breaking induced by the FI-term, in
the form
\be
G^{(1,1)}_\xi(1|2)=-\frac{(D^+_{\1})^4(D^+_{\2})^4}{ \bo_{1\xi}}
\delta^{12}(X_{\s 1}-X_{\s 2})
\exp (v_{2\xi}-v_{1\xi}) {1\over (u^+_1u^+_2)^3}~.
\label{E20}
\ee
Using eq.~\g{propbo}, one can check that this Green function satisfies 
the defining equation 
\be
{\cal D}^{\p}_{\xi\;1}G^{(1,1)}_\xi(1|2) = \delta_{\A}^{(3,1)}(1|2)~.
\ee 

The covariantly-analytic representation for the harmonic superfields in the 
same background is given by
\be
\hat{V}^{\p}_\xi=
e^{v_\xi}\tilde{V}^{\p} e^{-v_\xi},\quad \hat{q}^+=e^{v_\xi} q^+~.
\label{E21}
\ee
When using the same relations as in sect.~2, one can rewrite the modified 
$q^+$ propagator to the covariantly-analytic form as well. Its symmetry 
properties with respect to an exchange of the arguments become more 
transparent in the covariantly-analytic representation since 
the corresponding background harmonic derivatives are flat there by 
definition. We do not write down the explicit formulas here, since they 
can be easily obtained along the lines of sect.~2. 

Let us now briefly discuss how adding the above FI-term affects the 
quantum calculations of sect.~4. When using the modified hypermultiplet 
propagator (\ref{E20}) in perturbative calculations of the induced 
hypermultiplet self-interaction, one should consider a harmonic (one-loop) 
supergraph with an arbitrary (even) number of the external hypermultiplet 
legs, whose loop consists of both the hypermultiplet propagators 
and the $V^{\p}_\xi$ propagators in the alternative order. 
Since we can ignore all the extra terms with derivatives 
in the hypermultiplet
propagators, it is enough to concentrate on the $\theta$-dependent but 
otherwise constant factors. The gauge symmetry argument implies that the 
relevant terms contributing to the Wilsonian low-energy effective action 
have the structure
\be
\sum_n c_n \int d^{4}x d^8\theta du\, 
({\cal D}_{\xi}^{\m}\bar{q}^+ q^+)^{2n}\to
\int d^{4}x d^8\theta du\, f(v_{\xi}^{\m}\bar{q}^+ q^+) 
\label{E30}
\ee
to be determined by an even function $f$. The particular structure of the 
$\xi$-dependent terms with respect to the Grassmann coordinates 
in eqs.~\g{E6} and \g{E7} however implies that no $(\theta^-)^4$-terms can 
be formed out of them. In this sense, our induced (Taub-NUT) solution is 
stable against adding the FI-term.
\vglue.2in

\setcounter{equation}{0}
\section{Beyond perturbation theory}

\hspace{0.5cm} 
Since $N=2$ supersymmetry severely restricts the field couplings, it is quite 
natural to ask about other possible ways of generating a non-trivial 
hypermultiplet self-interaction provided that $N=2$ supersymmetry of the 
effective Lagrangian is maintained. As long as the $SU(2)_A$ automorphism 
symmetry of the $N=2$ supersymmetry algebra is not broken, it is clear that 
the perturbatively-induced Taub-NUT self-interaction (whose hyper-K\"ahler 
potential does not depend upon harmonic variables explicitly) is the 
{\it only~} non-trivial solution for a charged $N=2$ matter. Therefore, 
we should break this global symmetry in our $N=2$ supersymmetric gauge 
theory in some {\it controlled~} way, i.e. while maintaining {\it another} 
global symmetry of the free hypermultiplet action, in order to be able to 
make a concrete proposal.

The gauging procedure (or the coset construction) is known to be a powerful
method for generating new hyper-K\"ahler metrics (see e.g., ref.~\cite{AP}). 
In the harmonic superspace,
the additional resources for generating new hyper-K\"ahler potentials
${\cal L}^{(+4)}_{\rm ind.~int.}$ are given by (i) gauging isometries of the
moduli space of hypermultiplets, and (ii) adding (electric) FI terms. For
example, given two hypermultiplets $q^+_A\in \underline{2}$ of $SU(2)_f$, 
one can gauge a $U(1)$ subgroup of $SU(2)_f$ and simultaneously add 
a FI term as follows~\cite{GIOT}:
\be
S_{EH}=-\fracmm{1}{2\kappa^2}
\int d\zeta^{(-4)} \left[ q^{a+}_A D^{\p}q^+_{aA}+
V^{\p}_L\left(\frac{1}{2}\varepsilon^{AB}q^{a+}_Aq^+_{Ba}+\xi^{\p}\right)
\right]~,\label{F1}
\ee
where $V^{\p}_L$ is an $N=2$ gauge potential without a kinetic term
(Lagrange multiplier), and the $\xi^{\p}=u^+_iu^+_j\xi^{(ij)}$ was 
introduced in the previous sect.~5. The action $S_{EH}$ of eq.~(\ref{F1}) 
has the manifest (Pauli-G\"ursey) global $SU(2)_{PG}$ symmetry, which 
rotates the lower-case Latin indices only. At the same time, the $N=2$ SUSY 
automorphisms $SU(2)_{\rm A}$ are explicitly broken down to 
an $U(1)$ symmetry due to the presence of the charged `constant' 
$~\xi^{\p}\,$.

Eq.~(\ref{F1}) can be rewritten after some algebra to the form~\cite{GIOT}
\be
S_{EH}=-\fracmm{1}{2\kappa^2}
\int d\zeta^{(-4)} \left[  q^{a+} D^{\p}q^+_{a} -
\fracmm{(\xi^{\p})^2}{(q^{a+}u^-_{a})^2}\right]~. \label{F2}
\ee
It is clear from eq.~(\ref{F2}) that $\VEV{q^{a+}}\neq 0$ 
(the Higgs branch~!), just for the action $S_{EH}$ to make sense. 
Eq.~(\ref{F2}) takes the particularly simple form in 
terms of a real $\omega$-hypermultiplet which is {\it dual} to the 
$q^+$-hypermultiplet~\cite{GI2}. After the change of variables, 
$q^+_i=u^+_i\omega + u^-_if^{\p}$,
with an analytic (Lagrange multiplier) superfield $f^{\p}$, 
one finds~\cite{GIOT}
\be
S_{EH}=-\fracmm{1}{2\kappa^2}
\int d\zeta^{(-4)} \left[ \left(D^{\p}\omega\right)^2-
\fracmm{(\xi^{\p})^2}{\omega^2}\right]~.\label{F3}
\ee
The action $S_{EH}$ has the following $SU(2)_B$ symmetry~\cite{GIOT}:
\be
\delta\omega =c^{\m}D^{\p}\omega - c^{\s +-}\omega~,\label{F4}
\ee
where $c^{\m}=c^{ij}u^-_iu^-_j$ and  $c^{\s +-}=c^{ij}u^+_iu^-_j$ 
are the infinitesimal parameters of the $SU(2)_B$. 
This symmetry is basically the modified form of the $SU(2)_{PG}$ symmetry 
of the action \g{F1}. The automorphism $SU(2)_A$ group in the $\omega$ 
representation is given by the diagonal $SU(2)$ subgroup in the product of 
$SU(2)_{PG}$ and the automorphism group in the $q$ representation. 
This $SU(2)_A$ is, however, broken to its $U(1)$ subgroup in eq.~(\ref{F3}). 
Therefore, in any case, the action $S_{EH}$ has the $U(2)=SU(2)\times U(1)$ 
isometry group and, in its component form, it is known to lead to the 
four-dimensional {\it Eguchi-Hanson} gravitational instanton target 
metric~\cite{eh} for the physical bosonic fields of the hypermultiplet, 
in the form suggested in ref.~\cite{AGF}. Eq.~(\ref{F1}) can be easily 
generalized further, to the case of several hypermultiplets in 
$\underline{N_f}$ of $SU(N_f)$ whose Cartan generators are gauged in a 
presence of the $(N_f-1)$ FI terms~\cite{GIOT}. Then one gets the higher 
dimensional (Calabi) generalizations of the Eguchi-Hanson 
metric~\cite{Calabi}.

In order to address the very interesting question of a {\it dynamical} 
generation of the Eguchi-Hanson hyper-K\"ahler potential in eq.~(\ref{F3}), 
we would like to stress that the $SU(2)_B$ symmetry (\ref{F4}) is enough
to fix the form of the EH-potential {\it completely}. Also note that 
$SU(2)_B$ is just a subgroup of the whole symmetry group of the free $\omega$ 
hypermultiplet action, which includes, in particular, the whole $N=2$ 
superconformal group $SU(2,2|2)$. In other words, the EH-potential seems to 
be the {\it only~} deformation of the free hypermultiplet action that is 
consistent with the extra $SU(2)_B$ global symmetry when the automorphism 
symmetry $SU(2)_A$ is broken to its $U(1)_A$ subgroup by the charged 
constant $\xi^{\p}$. Accordingly, the dynamical mechanism we are looking for, 
should accomplish just two things: first, partially break the internal 
symmetry of the free hypermultiplet action down to $N=2$ supersymmetry 
{\it and~} $SU(2)_B\neq SU(2)_A$ and, second, break $SU(2)_A$ down to its 
$U(1)_A$ subgroup. 

These goals can be achieved in a nice geometrical way when using the recently
proposed {\it brane technology}~\cite{HWi,Wi}, where the quantum 
four-dimensional $N=2$ supersymmetric gauge theories are considered in the 
common world-volume of the M-theory/type-IIA superstring {\it branes}. 
The relevant BPS-like brane configuration should consist of the solitonic 
5-branes, Dirichlet 4- and 6-branes (in the type IIA picture) which are 
{\it intersecting at angles}~\cite{two} instead of being parallel as in 
ref.~\cite{Wi}. The vector distance between the two 5-branes along some three 
hidden ten-dimensional spacetime directions can then be identified with the FI 
parameter $\xi^{(ij)}$, whereas the intersecting angles of the D-4-branes 
can be identified with the constant gauge field strength 
components of the t'Hooft {\it toron}~\cite{Ht} after replacing 
the uncompactified four-dimensional common brane world-volume 
(i.e. the macroscopic space-time) by a torus, and imposing 
the twisted boundary conditions as in ref.~\cite{Ht}.

In the context of the $SU(2)$ gauge field theory, torons can be defined as 
the solitonic solutions having topological charge $Q=\frac{1}{2}$, i.e. as 
the {\it fractionally} charged instantons. The classical toron solution takes 
the form of an {\it analytic} (a half of supersymmetries is preserved~!) 
gauge superfield $V^{\p}_{\rm toron}$ ({\it cf.} ref.~\cite{yung}). 
It has an extra $SU(2)_{\rm toron}$ global symmetry given by the diagonal 
subgroup of $SU(2)_G\times SU(2)_R$, where the $SU(2)_G$ is the global 
part of gauge symmetry and $SU(2)_R\cong SO(3)$ is the rotational symmetry 
(i.e. the little group of the Lorentz symmetry). Unfortunately, within our 
pedestrian approach in this section, we are unable to prove that the 
$SU(2)_{\rm toron}$ does coincide with $SU(2)_{PG}$ or $SU(2)_B$ that are 
relevant for the induced Eguchi-Hanson self-interaction, though it seems us 
to be quite plausible. Up to this technical assumption, however, the net
impact on the allowed hypermultiplet self-interaction is just given by 
(i) a conservation of the extra $SU(2)_B\neq SU(2)_A$ global symmetry, and 
(ii) $\xi^{ij}\sim \delta^{ij}\exp(-2\pi^2/g^2)\neq 0$. As is well-known in 
the situation with {\it unbroken} $N=2$ SUSY, the Higgs branch receives 
neither scale- nor mass-dependent corrections~\cite{BF,APS}. The non-trivial 
(EH) hypermultiplet self-interaction therefore implies that the $N=2$ SUSY 
has to be spontaneously broken. It may be worthy to investigate this issue
in the framework of spontaneously broken $4D$, $N=2$ 
supergravity~\footnote{See ref.~\cite{IT} for a recent discussion
of the spontaneously broken $4D$, $N=2$ supergravity.}.

The dynamically generated FI term and the toron solution are, in fact, 
related to each other by  $N=2$ supersymmetry. The usual instantons are known 
to be unable to generate a non-vanishing gluino condensate because of the 
well-known index theorem. Indeed, one needs at least four fermions in order 
to saturate a non-vanishing correlator in the instanton 
background~\cite{ITEF}. Unlike the instantons, the torons do generate the 
non-vanishing gluino condensate~\cite{Zy},
\be
 \VEV{\lambda^i\lambda^j}=\Lambda^3 (\xi^2)^{ij}~,\quad {\rm with}\quad
 (\xi^2)^{ij}\sim \delta^{ij}\exp\left(-\fracmm{4\pi^2}{g^2}\right)~,
\label{F5}
\ee
where $\lambda^i$ is the fermionic $N=2$ superpartner of the gauge field
(i.e. $N=2~$ gaugino). As is well-known also, the non-perturbatively-generated 
gluino condensate has to be space-time {\it independent} due to the 
supersymmetric Ward identities~\cite{ITEF}. It just guarantees the constancy 
of $\xi^{ij}$. The constant parameter $\xi^{ij}$ entering eq.~(\ref{F5}) can 
now be identified with the FI constant. 

A combination of the perturbatively generated Taub-NUT-type self-interaction 
with the (presumably) non-perturbatively generated Eguchi-Hanson-type 
self-interaction could lead to the hyper-K\"ahler potential just given by a 
sum of the two potentials (in the {\it mixed} Coulomb-Higgs branch, if any).  
The corresponding hyper-K\"ahler metric is neither the Taub-NUT metric nor 
the Eguchi-Hanson one, but it reduces to them in particular limits.
\vglue.2in

\section*{Acknowledgements}

E.A.I. and B.M.Z. acknowledge a partial support from the Russian 
Foundation of Basic Research (RFBR) and the 
`Deutsche Forschungsgemeinschaft' 
(DFG) under the Projects RFBR No. 96--02--17634 and RFBR-DFG No. 
96--02--00180, respectively, the INTAS under the Projects Nos. 93--127 and 
94--2317, the Dutch NWO grant and the grant under the Heisenberg-Landau 
Program. They are grateful to Profs. 
O. Lechtenfeld and D. L\"ust for useful discussions and a kind 
hospitality at the Hannover and Humboldt Universities
where a part of this work was done. S.V.K. acknowledges the support 
of the `Deutsche Forschungsgemeinschaft' and the NATO grant CRG 930789.
He would like to thank the Bogoliubov Laboratory of Theoretical Physics 
in Dubna and the Theory Division of CERN, where some parts of this work 
were completed, for a kind hospitality extended to him.
\vglue.2in

\newpage

\section*{Appendix: from HSS to component results}

In this Appendix we derive the induced Taub-NUT metric and the 
induced scalar
potential in the presence of central charges from HSS, 
following the lines of ref.~\cite{GIOS}.

The general procedure to get the component form of the bosonic non-linear
sigma-model from a hypermultiplet action in HSS consists of the following 
steps:
\begin{itemize}
\item expand the equations of motion in Grassmann variables, and ignore all
the fermionic field components,
\item solve the kinematical differential equations (on the sphere
$S^2\sim SU(2)/U(1)$) for the auxiliary field components, thus eliminating 
the infinite tower of them in the harmonic expansion of the hypermultiplet
HSS superfields,
\item substitute the solution into the original 
hypermultiplet action in HSS,
and integrate over all the anticommuting and harmonic coordinates.
\end{itemize}
In our case, with the HSS hypermultiplet action
$$ S_{TN} =-\int d\zeta^{(-4)}\left[ \bar{q}^+{\cal D}^{\p}q^+ 
+\fracmm{\lambda}{2}
(\bar{q}^+)^2(q^+)^2\right]~,\eqno(A.1)$$
the HSS equations of motion in the analytic $N=2$ superspace
$\zeta=\{x^{m}_{\rm A},\theta^+_{\alpha},
\bar{\theta}^+_{\dot{\alpha}},u^{\pm}_i\}$ for the analytic superfield
$q^+(\zeta)$  are given by
$$ {\cal D}^{\p}q^+ +\lambda(\bar{q}^+q^+)q^+=0~,\eqno(A.2)$$
where the analytic harmonic derivative ${\cal D}^{\p}$ is
$${\cal D}^{\p} = \partial^{\p}
-2i\theta^+\sigma^m\bar{\theta}^+\partial_m
+i(\theta^+)^2\bar{Z}+i(\bar{\theta}^+)^2Z~. 
\eqno(A.3)$$
The bosonic terms in the $\theta$-expansion of $q^+$ read
$$\eqalign{
q^+(\zeta)= & F^+(x_{\rm A},u)+i\theta^+\sigma^m\bar{\theta}^+A_m^-
(x_{\rm A},u)+\theta^+\theta^+M^-(x_{\rm A},u)\cr
& +\bar{\theta}^+\bar{\theta}^+ N^-(x_{\rm A},u)+\theta^+\theta^+
\bar{\theta}^+\bar{\theta}^+P^{(-3)}(x_{\rm A},u)~.\cr}\eqno(A.4)$$

In a presence of central charges, some of the kinematical equations of 
motion in the $(x_{\rm A},u)$ space get modified ({\it cf.} 
ref.~\cite{GIOS}),
$$\eqalign{
\partial^{\p} F^+= & -\lambda(\bar{F}^+F^+)F^+~,\cr
\partial^{\p} A^-_m= & 2\partial_mF^+
-\lambda(\bar{F}^+F^+)A^-_m  -\lambda(F^+)^2\bar{A}^-_m~,\cr
\partial^{\p} M^{-}= & -\lambda(F^+)^2\bar{N}^-
-2\lambda(\bar{F}^+F^+)M^- -i\bar{Z}F^+~,\cr
\partial^{\p} N^- = & -\lambda(F^+)^2\bar{M}^-
-2\lambda(\bar{F}^+F^+)N^- -iZF^+ ~.\cr}
\eqno(A.5)$$

After integrating over the Grassmann variables in the action (A.1) 
and using
the kinematical equations of motion, one finds that the bosonic 
action reduces to
$$
S_{\rm B}= \frac{1}{2}\int d^4xdu
\left[ A_m^-\partial^m\bar{F}^+ - \bar{A}_m^-\partial^mF^+
- i(\bar{N}^- Z + \bar{M}^- \bar{Z}) F^+ - i\bar{F}^+( Z M^- 
+ \bar{Z} N^-) \right]~.\eqno(A.6)$$
The kinematical equations for $F^+$ and $A^-_m$ can be easily solved, as in
ref.~\cite{GIOS}. Unlike that in ref.~\cite{GIOS}, the auxiliary 
fields $M^-$ and $N^-$ at $Z\neq 0$ now contribute too. Using the 
convenient parametrization~\cite{GIOS}
$$ F^+(x,u)=f^i(x)u^+_i\exp\left[\lambda f^{(j}(x)\bar{f}^{k)}(x)u_j^+u^-_k
\right]~,\eqno(A.7)$$
one finds that
$$ S_{\rm B}= \int d^4x\,\left\{
g_{ij}\partial_mf^i\partial^mf^j+\bar{g}^{ij}\partial_m\bar{f}_i\partial^m
\bar{f}_j + h^i{}_j\partial_mf^j\partial^m\bar{f}_i -V(f)\right\}~,
\eqno(A.8)$$
where the metric is given by~\cite{GIOS}
$$g_{ij}=\fracmm{\lambda(2+\lambda f\bar{f})}{4(1+\lambda f\bar{f})}
\bar{f}_i
\bar{f}_j~,\quad
\bar{g}^{ij}=\fracmm{\lambda(2+\lambda f\bar{f})}{4(1+\lambda f\bar{f})}
f^if^j~,$$
$$ h^i{}_j=\delta^i{}_j(1+\lambda f\bar{f})
-\fracmm{\lambda(2+\lambda f\bar{f})}{2(1+\lambda f\bar{f})}f^i\bar{f}_j~,
\quad
{\rm and}\quad f\bar{f}\equiv f^i\bar{f}_i~.\eqno(A.9)$$
The metric (A.9) takes the standard Taub-NUT form~\cite{EGH}
$$ ds^2=\fracmm{r+M}{2(r-M)}dr^2+\frac{1}{2}(r^2-M^2)
(d\vartheta^2+\sin^2\vartheta d\varphi^2) +2M^2
\left(\fracmm{r-M}{r+M}\right)
(d\psi +\cos\vartheta d\varphi)^2~,\eqno(A.10)$$
after the change of variables
~\cite{GIOS}
$$\eqalign{
f^1=& \sqrt{2M(r-M)}\cos\fracmm{\vartheta}{2}
\exp\fracmm{i}{2}(\psi +\varphi)~,\cr
f^2=& \sqrt{2M(r-M)}\sin\fracmm{\vartheta}{2}
\exp\fracmm{i}{2}(\psi -\varphi)~,\cr}
\eqno(A.11)$$
with
$$
f\bar f = 2M(r-M)~, \quad
r\geq M\equiv \fracmm{1}{2\sqrt{\lambda}}~,\eqno(A.12)$$
and $M$ being the mass of the Taub-NUT gravitational instanton. 
The non-vanishing auxiliary fields $M^-$ and $N^-$ lead, in addition, 
to the non-trivial scalar potential,
$$ V(f)=\abs{Z}^2\fracmm{f\bar{f}}{1+ \lambda f\bar{f}}~.
\eqno(A.13)$$
It is the last equation (A.13) that was quoted in the text 
(see eq.~\g{D22} in sect.~4). It is also worth noticing that the quadratic 
terms in eq.~(A.13) are $\lambda$-independent as expected, 
since the BPS mass 
$m^2=\abs{Z}^2$ is protected by supersymmetry and, therefore, 
it is not going to be renormalized. 
\vglue.2in


\newpage

\begin{thebibliography}{99}

\bibitem{SW} N. Seiberg and E. Witten, Nucl. Phys. B426 (1994) 19, 
{\it ibid.}
B431 (1994) 484.
\bibitem{GIK1} A. Galperin, E. Ivanov, S. Kalitzin, V. Ogievetsky and
               E. Sokatchev. Class. Quantum Grav. 1 (1984) 469.
\bibitem{GI2} A. Galperin, E. Ivanov,  V. Ogievetsky and
             E. Sokatchev. Class. Quantum Grav. 2 (1985) 601, 617.
\bibitem{BBIKO}  I. Buchbinder, E. Buchbinder, E. Ivanov, S. Kuzenko and
B. Ovrut, {\it Effective action of the N=2 Maxwell multiplet in harmonic
superspace}, Princeton, Pennsylvania, Dubna and Hannover preprint,
IASSNS-HEP-97-6, UPR-733T, JINR E2-97-82 and ITP-UH-09/97; hep-th/9703147.
\bibitem{BBKO} I. Buchbinder, E. Buchbinder, S. Kuzenko and B. Ovrut, 
{\it The background field method for N=2 super-Yang-Mills 
theories in harmonic
superspace}, Princeton, Pennsylvania and Hannover preprint, 
IASSNS-HEP-97-32T,
 UPR-745T, and ITP-UH-12/97; hep-th/9704214.
\bibitem{SS}  J. Scherk and J. Schwarz, Nucl. Phys. B153 (1979) 61.
\bibitem{Seib} N. Seiberg, Phys. Lett. 206B (1988) 75.
\bibitem{Ke} S. Ketov, {\it The effective hyper-K\"ahler potential 
in the N=2 supersymmetric QCD}, DESY and Hannover preprint, DESY 97-007
and  ITP-UH-03/97; hep-th/9701158.
\bibitem{BF} R. Barbieri, S. Ferrara, L. Maiani, F. Palumbo and C. Savoy, 
Phys. Lett. 115B (1982) 212.
\bibitem{APS} P. Argyres, M. Plesser, N. Seiberg and E. Witten, 
Nucl. Phys. B461 (1996) 71.
\bibitem{GIOS} A. Galperin, E. Ivanov, V. Ogievetsky and E. Sokatchev, 
Commun. Math. Phys. 103 (1986) 515.
\bibitem{GIOT} A. Galperin, E. Ivanov, V. Ogievetsky and P. Townsend. 
Class. Quantum Grav. 3 (1986) 625.
\bibitem{Z1} B. Zupnik, Sov. J. Nucl. Phys. 44 (1986) 512;\\
             in the Proceedings of the 3rd Intern. Seminar
``Group-Theoretical Methods in Physics'', Yurmala, USSR, 1985;
Nauka Publishers, Moscow, 1986, Vol. 1, p. 52.
\bibitem{BSS} L. Brink, J. Schwarz and J. Scherk, 
Nucl. Phys. B121 (1977) 77.
\bibitem{Fa} P. Fayet. Nucl. Phys. B 263 (1986) 649.
\bibitem{Ku} S. Kuzenko, {\it The off-shell massive 
hypermultiplets revisited},
Hannover preprint ITP-UH-13/97; hep-th/9704002.
\bibitem{CWZ} S. Coleman, J. Wess and B. Zumino, 
Phys. Rev. 177 (1969) 2239.
\bibitem{AFp} L. Alvarez-Gaum\'e and D. Freedman, 
Commun. Math. Phys. 91 (1983) 87.
\bibitem{IZ} E. Ivanov and B. Zupnik, in preparation.
\bibitem{Fa2} P. Fayet. Nucl. Phys. B113 (1976) 135.
\bibitem{Ka} A. Kapustnikov. Yadern. Fiz. 45 (1987) 275.
\bibitem{ZT} B. Zupnik and L. Tolstonog, Sov. J. Nucl. Phys. 46 (1988) 160.
\bibitem{AP} U. Lindstr\"om and M. Ro\v{c}ek, Nucl. Phys. B222 (1983) 285;\\
N. Hitchin, A. Karlhede, U. Lindstr\"om and M. Ro\v{c}ek, Commun. Math. Phys.
108 (1987) 535; \\
I. Antoniadis and B. Pioline, {\it Higgs branch, hyper-K\"ahler quotient and 
duality in SUSY N=2 Yang-Mills theories}, Palaiseau preprint CPTH-S459.0796; 
hep-th/9607058.
\bibitem{eh} T. Eguchi and A. J. Hanson, Phys. Lett. 74B (1978) 249.
\bibitem{AGF} Th. L. Curtright and D. Z. Freedman, Phys. Lett. 90B (1980) 
71;\\ 
L. Alvarez-Gaum\'e and D. Z. Freedman, Phys. Lett. 94B (1980) 171.
\bibitem{Calabi} E. Calabi, Ann. Sci. de l'E.N.S. 12 (1979) 266. 
\bibitem{HWi} A. Hanany and E. Witten, {\it Type IIB superstrings,
BPS monopoles and three-dimensional gauge dynamics}, Princeton preprint
IASSNS--HEP--96--121; hep-th/9611230.
\bibitem{Wi} E. Witten, {\it Solutions of four-dimensional field
theory via M theory}, Princeton preprint IASSNS-HEP-97-19;
hep-th/9703166.
\bibitem{two} M. Berkooz, M. R. Douglas and R. G. Leigh, Nucl. Phys. B480
(1996) 265; \\
S. V. Ketov, in preparation.
\bibitem{Ht} G. t'Hooft, Commun. Math. Phys. 81 (1981) 267. 
\bibitem{yung} A. Yung, {\it Instanton-induced effective Lagrangian in the
Seiberg-Witten model}, Swansea preprint SWAT/96/111; hep-th/9605096.
\bibitem{IT} L. Girardello, M. Porrati and A. Zaffaroni, 
{\it Spontaneously broken N=2 supergravity without light mirror fermions}, 
New-York, Princeton and Milano preprint, NYU-TH-97/03/02, IASSNS-HEP-97/25 
and IFUM/FT-558; hep-th/9704163.
\bibitem{ITEF} V. Novikov, M. Shifman, A. Vainstein, and V. Zakharov,
Nucl. Phys. B229 (1983) 381, 394, 407.
\bibitem{Zy} A. Zhitnitsky, Nucl.Phys. B340 (1990) 56.
\bibitem{EGH} T. Eguchi, P. Gilkey and A. Hanson, Phys. Rep. 66 (1980) 213.
\end{thebibliography}
\end{document}